%% file: main.tex
\documentclass[a4paper,12pt,parskip=half,english,numbers=noenddot]{scrartcl}

\include{header}

\begin{document}

\input{sections/summery}
\input{sections/introduction}
\input{sections/basics}
\input{sections/multiscale}
\input{sections/discretization}
\input{sections/numerics}

\input{sections/conclusions}

%\appendix
%\section*{Appendix}\label{sec:appendix}
%\input{sections/sec_appendix}

%------------------------------------------------------------------
% Bibliography
%------------------------------------------------------------------	
	
\bibliographystyle{plain}
\bibliography{literature}

\end{document}

%% file: header.tex
\usepackage{babel}
\usepackage[T1]{fontenc}
\usepackage{lmodern,textcomp,latexsym}
\usepackage{amsthm,amssymb,amsmath}
\usepackage{graphicx,psfrag,color,rotate}
\usepackage{paralist,threeparttable}
\usepackage[footnotesize,labelfont={bf}]{caption}
\usepackage{alphalph}
\usepackage{longtable,booktabs,multirow}
\usepackage{tabularx,nth}
\usepackage{tikz}
\usepackage{pifont}
\usepackage{enumerate}
\usepackage{ulem,upgreek}
\usepackage{cancel}

\usepackage{float}

\usepackage{hyperref}
\usepackage[boxed]{algorithm2e}

\usepackage{verbatim}

\DeclareCaptionStyle{table_new}{font={footnotesize},justification=raggedright}
\setdefaultenum{1)}{\theenumi.1)}{}{}

\newtheorem{remark}{Remark}

\renewcommand{\vec}{\boldsymbol}
\renewcommand{\d}[1][]{\ensuremath{\,\mathrm{d}#1}}

\newcommand{\indi}[1]{\ensuremath{\operatorname{#1}}}
\newcommand{\op}[1]{\ensuremath{\operatorname{#1}}}

\newcommand{\nablaq}[1]{\ensuremath{\nabla_{\hspace{-0.9mm}\mbox{\begin{scriptsize}$\mathrm{#1}$\end{scriptsize}}}}\hspace{0.3mm}}

\makeatletter
\renewcommand*{\env@matrix}[1][*\c@MaxMatrixCols c]{%
  \hskip -\arraycolsep
  \let\@ifnextchar\new@ifnextchar
  \array{#1}}
\makeatother

\newcommand{\s}{{\mathfrak{s}}}
\newcommand{\ds}{\ensuremath{\tilde{d}}}

\graphicspath{{./pictures/}}

\DeclareOldFontCommand{\it}{\normalfont\itshape}{\mathit}

\renewcommand{\d}{\operatorname{d}\!}

\newcommand{\RVE}{\ensuremath{\Omega_0}}

\makeatletter
\DeclareRobustCommand\bigop[1]{%
  \mathop{\vphantom{\sum}\mathpalette\bigop@{#1}}\slimits@
}
\newcommand{\bigop@}[2]{%
  \vcenter{%
    \sbox\z@{$#1\sum$}%
    \hbox{\resizebox{\ifx#1\displaystyle.9\fi\dimexpr\ht\z@+\dp\z@}{!}{$\m@th#2$}}%
  }%
}
\makeatother

\renewcommand{\vec}[1]{\pmb{#1}}

\usepackage{tikz}
\usetikzlibrary{decorations.pathmorphing} % for snake lines
\usetikzlibrary{matrix} % for block alignment
\usetikzlibrary{arrows, arrows.meta} % for arrow heads
\usetikzlibrary{calc} % for manimulation of coordinates
\usetikzlibrary{shapes}
\usetikzlibrary{shapes.geometric}

\usepgfmodule{nonlineartransformations}
\usepgflibrary{curvilinear}

\usepackage{tikz-3dplot} % needs tikz-3dplot.sty in same folder

\usepackage{bm}

\tikzstyle{block} = [draw,rectangle,thick,minimum height=2em,minimum width=2em]
\tikzstyle{sum} = [draw,circle,inner sep=0mm,minimum size=2mm]
\tikzstyle{connector} = [->,thick]
\tikzstyle{line} = [thick]
\tikzstyle{branch} = [circle,inner sep=0pt,minimum size=1mm,fill=black,draw=black]
\tikzstyle{guide} = []
\tikzstyle{snakeline} = [connector, decorate, decoration={pre length=0.2cm,
	post length=0.2cm, snake, amplitude=.4mm,
	segment length=2mm},thick, magenta, ->]

%% file: sections/summery.tex
\begin{center}
\large{\textbf{Computational homogenization of phase-field fracture.}}

{\large Felix Schmidt$^{a}$, Stefan Schu{\ss}$^{a}$ and Christian Hesch$^{a}$\footnote{Corresponding author. E-mail address: christian.hesch@uni-siegen.de}}

{\small
\(^a\) Chair of Computational Mechanics, University of Siegen, Germany
}

\end{center}

\vspace*{-0.1cm}\textbf{Abstract}

\textbf{Abstract.} 
In this contribution we investigate the application of phase-field fracture models on non-linear multiscale computational homogenization schemes. In particular, we introduce different phase-fields on a two-scale problem and develop a thermodynamically consistent model. This allows on the one hand for the prediction of local micro-fracture patterns, which effectively acts as an anisotropic damage model on the macroscale.  On the other and, the macro-fracture phase-field model allows to predict complex fracture pattern with regard to local microstructures.  Both phase-fields are introduced in a common framework, such that a joint consistent linearization for the Newton-Raphson iteration can be developed.  Finally, the limits of both models as well as the applicability are shown in different numerical examples. 


%% file: sections/introduction.tex
\section{Introduction}
The prediction of the overall mechanical response of materials with dedicated microstructures including various failure mechanisms require direct numerical simulations. However, fully resolved microstructures may lead to economically or technically unsustainable computational costs.  Therefore, it is convenient to apply computational homogenization techniques, assuming that the material may appear perfectly homogeneous on a higher, macroscale and highly heterogeneous at a lower, microscale.  To be more specific, we assume that the scale separation is valid, i.e.\ that the wavelengths of physical fields at the higher scale are very much larger than the dimensions of heterogeneities at the lower scale, see \cite{sanchez-palencia_homogenization_1983}. To take microstructural information about the morphology and material properties into account,  various homogenization methods have been introduced, with focus here on computational homogenization like FE$^2$ and IGA$^2$.

In the context of first-order homogenization schemes, we refer to \cite{hill_elastic_1952,hashin_variational_1963,hill_self-consistent_1965,mori_average_1973,willis_bounds_1977} for fundamental analytical approaches and to \cite{smit_prediction_1998,miehe_computational_1999-1,feyel_fe2_2000,kouznetsova_approach_2001} for seminal contributions to two-scale finite-element (FE) simulations. In the context of higher-order and generalized continua, analytical approaches have been explored by \cite{diener_bounds_1984,gambin_higher-order_1989,boutin_microstructural_1996,drugan_micromechanics-based_1996},  whereas a general framework for a numerical two-scale analysis using materials up to the third order has been introduced in Schmidt et al. \cite{hesch2022}. 

Moreover, the numerical prediction of the initiation and propagation of brittle and ductile fracture pattern has gained tremendous attention in the past decade.  Griffith and Irwin \cite{griffith1921,irwin1958} formulated the propagation of brittle fracture by assuming that the material fails locally upon the attainment of a specific fracture energy related to a critical energy-release rate.  A straightforward minimisation of the fracture energy is not feasible due to the arising discontinuities.  Therefore, diffusive crack zones  have been introduced to avoid the modeling of discontinuities. In particular, phase-field methods provide a framework to formulate variational fracture models based on regularized energy functionals embedded within a global minimizer, cf. Francfort and Marigo \cite{FrancfortMarigo1998}.

Nowadays, phase-field methodologies have been applied to a wide range of brittle and ductile fracture including multiphysical environments, see, among many other, Miehe et al.\ \cite{miehe2010d}, Hesch et al.\ \cite{hesch2014b,hesch2016b}, Borden et al.\ \cite{borden2014}, Kuhn et al.\ \cite{kuhn2015}, Verhoosel and De Borst \cite{verhoosel2013}, Paggi and Reinoso \cite{paggi2017b}, Teichtmeister et al.\ \cite{teichtmeister2017}, Zhang et al.\ \cite{zhang2017}, Dittmann et al.\ \cite{dittmann2017a,dittmann2018b}, Heider and Markert \cite{heider2017}, Bryant and Sun \cite{bryant2018} and Aldakheel et al.\ \cite{aldakheel2018} for brittle fracture and Dittmann et al. \cite{dittmann2018} and Schulte et al. \cite{schulte_isogeometric_2020} on ductile fracture.

This contribution is devoted to the application of phase-field methods on multiscale computational homogenization schemes. Obviously, a phase-field fracture approach can be applied to both scales on a two-scale homogenization method, leading to two different dissipation functions. We aim at a thermodynamical consistent formulation to avoid unphysical energy transfer either in the wrong direction or between the different dissipation potentials. This issue has already been investigated in the context of ductile fracture, using a multiplicative triple split of the deformation gradient to satisfy the Clausius-Planck inequality equation.  Analoguesly, we introduce here a triple split into local micro-fracture and global macro-fracture. For the former part, the model is restricted by the effective scale separation, leading to limitations in the applicability due to the absence of non-local and gradient contributions on the macroscale. Finally, we develop in the discrete setting a consistent linearization for the non-linear two-field two-scale problem. 

The manuscript is organized as follows. Seciton \ref{sec:gov} presents the basic governing equations for non-linear phase-field brittle fracture on a single scale. In Section \ref{sec:multscale}, the multiscale fracture formulation is presented, followed by the discretization and linearization in Section \ref{sec:disc}. Numerical examples are given in Section \ref{sec:numerics} and conclusions are drawn in \ref{sec:con}.

%% file: sections/basics.tex
\section{Governing equations}\label{sec:gov}

We start with a short summary of the governing equations for the macro phase-field fracture formulation at hand. Consider a Lipschitz bounded continuum body in its reference configuration \(\mathcal{B}_{0}\subset\mathbb{R}^{n}\) with \(n\in \{2,3\}\) undergoing a motion characterized by the deformation mapping
\begin{equation}
\vec{\varphi}(\vec{X}):\mathcal{B}_{0} \rightarrow \mathbb{R}^d  
\quad \mbox{such that} \quad\vec{x} = \vec{\varphi}(\vec{X}),
\end{equation}
which maps material points $\vec{X} \in \mathcal{B}_{0}$ of the reference configuration $\mathcal{B}_{0}$ onto points $\vec{x} \in \mathcal{B}_{t}$ of the current configuration $\mathcal{B}_{t} = \vec{\varphi}(\mathcal{B}_{0})$. The material deformation gradient is defined by $\vec{F} := \nablaq{\vec{X}}(\vec{\varphi}(\vec{X}))$ with $J \!:=\!\mbox{det}[\vec{F}]>0$. Next, a phase-field is introduced as
\begin{equation} 
\s(\vec{X}):\mathcal{B}_{0}\times\mathcal{T}\rightarrow [0,1]\subset\mathbb{R},\quad \mbox{with} \quad\dot \s \ge 0,
\end{equation}
where the value \(\s = 0\) refers to the undamaged and \(\s = 1\) to the fully broken state of the material.  Here, $(\dot{\bullet})$ refers to the derivative in time, such that $\dot \s \ge 0$ prevents healing effects. For the static system considered here, we re-interpret the temporal derivative by preventing healing from load step to load step. 

\subsection{Kinematics and regularised crack phase-field topology}

In analogy to the multiplicative decomposition of non-linear plasticity (cf.\ \cite{dittmann2018}), we define the elastic, fracture insensitive part of the deformation gradient, written in terms of principal stretches $\lambda_a$, as follows
\begin{equation}\label{eq:kinematic}
\vec{F}^\mathrm{e} = \sum\limits_{a=1}^d(\lambda_a^+)^{g(\s)}(\lambda_a^-)\vec{\mathfrak{n}}_a\otimes\vec{\mathfrak{N}}_a,
\end{equation}
postulating that fracture requires a local state of tension.  Here, \(\vec{\mathfrak{n}}_a\) and \(\vec{\mathfrak{N}}_a\) denote the principal directions of the left and right stretch tensors and $g(\s)$ a suitable degeneration function; in the most simplest setting $g(\s) = 1-\s$. Moreover, \(\lambda_a\) are principal stretches, decomposed into tensile and compressive components using \(\lambda_a^{\pm} = [(\lambda_a-1)\pm |\lambda_a-1|]/2+1\). Cracks can be considered as propagating internal boundaries \(\Gamma^{cr}(t)\subset\mathbb{R}^{n-1}\) and we assume that crack initiates or continues upon attainment of a critical local fracture energy density \(g_c\). Thus, the total energy within the sharp crack interface reads
\begin{equation}
E^{cr} = \int\limits_{\Gamma^{cr}}g_c\d\Gamma.
\end{equation}
Since the numerical evaluation of this sharp crack interface is not suitable within a standard finite element framework, a regularised crack interface using a specific regularisation profile \(\gamma(\s)\) is introduced, such that the critical fracture energy is approximated by 
\begin{equation}
\int_{\Gamma^{cr}}g_c\d\Gamma \approx \int_{\mathcal{B}_0}g_c\gamma_l(\s)\d V. 
\end{equation}
The chosen regularisation profile for a second order phase-field is assumed to take the form \(\s(x) = e^{\frac{-|x|}{2\,l}}\) (see Miehe et al. \cite{miehe2010,miehe2010d} for details),  which is the solution of the partial differential equation
\begin{align}\label{DGLs}
\s - l^2{\Delta\s} & = 0\;\text{in}\;\mathcal{B}_{0},\\
\nabla \s\cdot\vec{n} &= 0\;\text{on}\;\partial\mathcal{B}_{0}.
\end{align}
The corresponding variational formulation was first introduced by Francfort and Marigo \cite{FrancfortMarigo1998}. Therefore, we introduce the space of admissible solution an test functions for the phase field 
\begin{equation}
\mathcal{V}^{\s} = \{{\s}\in H^1(\mathcal{B}_0)|\dot{\s} \geq 0 \;\text{on}\; \mathcal{B}_0\},\quad\mathcal{V}^{\delta\s} = \{{\delta\s}\in H^1(\mathcal{B}_0)|{\delta\s} = 0 \;\text{on}\; \Gamma^{cr}\},
\end{equation}
where \(H^1\) denotes the Sobolev functional space of square integrable functions and derivatives. Applying integration by parts, the weak form of (\ref{DGLs}) can be written as
\begin{equation}
\int\limits_{\mathcal{B}_0} \frac{1}{l} {\delta\s}\, \s+l\nabla({\delta\s})\cdot\nabla(\s)\,\d V = 0.
\end{equation}
This, in turn, can be related to the regularized crack surface topology,
\begin{equation}\label{intgamma}
\Gamma_l^{cr}(\s) := \int\limits_{\mathcal{B}_0}\gamma_l(\s)\,\d V = \int\limits_{\mathcal{B}_0}\frac{1}{2l}\s^2+\frac{l}{2}\nabla(\s)\cdot\nabla(\s)\,\d V,
\end{equation}
such that the diffusive crack zone (\ref{intgamma}) converges to the sharp crack surface \(\Gamma^{cr}(t)\)  for \(l\rightarrow 0\).  

\begin{remark}
Often, a split in isochoric and volumetric parts using $\vec{F}^\mathrm{e} = J^{\frac{1}{n}}\bar{\vec{F}}^\mathrm{e}$ and $\lambda_a = J^{\frac{1}{n}}\bar{\lambda}_a$, is applied,  such that the isochoric elastic deformation gradient reads
\begin{equation}
\bar{\vec{F}}^{\mathrm{e}} = \sum\limits_{a} \left(\bar{\lambda}_a\right)^{g(\s)}\,\vec{n}_a\otimes\vec{N}_a\,,
\end{equation}
Moreover, assuming that fracture requires a local state of expansion, the fracture insensitive, volumetric part of the Jacobian is defined as
\begin{equation}
{J}^{\mathrm{e}}=\begin{cases}\left(J\right)^{g(\s)} & \text{if} \quad J  > 1\\ J & \text{else}\end{cases}\, .
\end{equation}
Depending on the specific situation,this approach has proven to be more robust. 
\end{remark}

\subsection{Balance equation}
Introducing the space of virtual or admissible test functions for the deformation 
\begin{equation}
\mathcal{V}^{\delta\varphi} = \{\delta\vec{\varphi}\in {H}^1(\mathcal{B}_0)\,|\,\delta\vec{\varphi}=\vec{0}\,\text{on}\,\Gamma^{\varphi}\},
\end{equation}
where $\Gamma^{\varphi}$ refers to the Dirichlet boundary,  the internal virtual work and the virtual work of the inertia terms along with the internal virtual work of the crack driving force reads
\begin{equation}\label{eq:internal}
\delta\Pi^{\mathrm{int},\varphi} = \int\limits_{\mathcal{B}_0}\vec{P}:\nablaq{\vec{X}}(\delta\vec{\varphi})\,\d V,\quad
\delta\Pi^{\mathrm{int},\s} = -\int\limits_{\mathcal{B}_0}\delta\s\,\mathcal{H}\,\d V.
\end{equation}
For localization, we apply integration by parts on \eqref{eq:internal}$_1$
\begin{equation}
\int\limits_{\mathcal{B}_0}\vec{P}:\nablaq{\vec{X}}(\delta\vec{\varphi})\,\d V = \int\limits_{\partial\mathcal{B}_0}\delta\vec{\varphi}\cdot\vec{P}\vec{N}\, \d A - \int\limits_{\mathcal{B}_0}\delta\vec{\varphi}\cdot(\nablaq{\vec{X}}\cdot \vec{P})\, \d V,
\end{equation}
where $\vec{N}$ is the outward normal in the reference configuration.  

Introducing $\vec{B}$ as local body load and $\vec{T}$ as local Neumann surfaces load on $\Gamma^\sigma$, noting that $\Gamma^\varphi\cap\Gamma^\sigma = \emptyset$ and $\Gamma^\varphi\cup\Gamma^\sigma = \Gamma$ with $\Gamma$ an open set on $\partial\mathcal{B}_0$,  we can equilibrate internal and external virtual work and obtain
\begin{equation}\label{eq:weakMech}
\int\limits_{\mathcal{B}_0}\vec{P}:\nablaq{\vec{X}}(\delta\vec{\varphi})\,\d V = \int\limits_{\mathcal{B}_0}\delta\vec{\varphi}\cdot\vec{B}\, \d V +
\int\limits_{\Gamma^\sigma} \delta\vec{\varphi}\cdot\vec{T}\, \d A.
\end{equation} 
In contrast, the virtual work of the phase-field driving force $\mathcal{H}$ is equilibrated with the regularised virtual work of the critical fracture energy
\begin{equation}\label{eq:weakPhase}
\int\limits_{\mathcal{B}_0}\delta\s\left(\mathcal{H}+\frac{g_c}{l}\s\right)+
g_c l\nabla(\delta\s)\cdot\nabla(\s)\,\d V = 0,
\end{equation}
such that we control the dissipated energy into the regularised crack interface. To demonstrate this, we set $\delta\vec{\varphi} = \dot{\vec{\varphi}}$ and $\delta\s = \dot{\s}$ and restate \eqref{eq:weakMech} as
\begin{equation}\label{eq:conti2}
\int\limits_{\mathcal{B}_0}\vec{P}:\,\dot{\vec{F}}\,\d V = 
\underbrace{\int\limits_{\mathcal{B}_0}\dot{\vec{\varphi}}\cdot\vec{B}\,\d V + \int\limits_{\Gamma^{\sigma}}\dot{\vec{\varphi}}\cdot\vec{T}\,\d A}_{P^{ext}}, 
\end{equation}
where $P^{ext}$ the external power.  Moreover, we obtain from \eqref{eq:weakPhase}
\begin{equation}\label{eq:contiPF}
\underbrace{\int\limits_{\mathcal{B}_0}\dot{\s}\frac{g_c}{l}\s+g_cl\nabla(\dot{\s})\cdot\nabla(\s)\,\d V}_{  \int\limits_{\mathcal{B}_0}g_c\dot{\gamma}_{l}\,\d V = D^{\indi{ma}}} =
-\int\limits_{\mathcal{B}_0}\mathcal{H}\dot{\s}\,\d V
\end{equation} 
such that
\begin{equation}\label{eq:1law}
 \underbrace{\int\limits_{\mathcal{B}_0}\vec{P}:\,\dot{\vec{F}} + \mathcal{H}\dot{\s}\,\d V}_{P^{\indi{int}}} + D^{\indi{ma}}  = P^{\indi{ext}}.
\end{equation}
which is the first law of thermodynamics excluding the kinetic energy.  Without proof,  we state that $\dot{\s} \geq 0$ ensures that the Clausius Duhem inequality equation is valid, see \cite{hesch2014} for details.  

%\begin{remark}
%Since a continuum theory can resolve every possible microstructure in a single scale by definition, a multiscale formulation can be applied in the discrete setting using a finite set of distinguishable material points for the calculation of the approximated field equations. Even for an artificial definition of a multiscale situation, the first order Taylor expansion of the microscale deformation map (introduced in the next Section) at two macroscale points $\vec{X}_1$ and $\vec{X}_2$, $\vec{X}_1 = \vec{X}_2 + \epsilon(\vec{X}-\vec{X}_2)$ with $\epsilon \rightarrow 0$, $\epsilon \neq 0$,  such that the microstructures overlap, may produce different states of deformation at the same physical position depending on the considered inhomogeneity.  
%\end{remark}

%% file: sections/multiscale.tex
\section{Multiscale phase-field boundary value problem}\label{sec:multscale}
Based on the previous considerations, we introduce often a phenomenological,  i.e.\ constitutive strain energy function,  such that the temporal derivative recovers the internal power $P^{\indi{int}}$ introduced in \eqref{eq:1law}.  However, the construction of phenomenological strain energy functions becomes nowadays more and more complex as production tools like additive manufacturing allow for (geometrically) inhomogeneous materials.  To capture these geometrical effects in a computationally efficient way, a dimensional expansion is introduced. In particular,  we introduce a representative volume element (RVE) $\Omega_0 = \Omega_0(\vec{X}) \subseteq\mathbb{R}^n$ to each point $\vec{X}\in\mathcal{B}_0\subset \mathbb{R}^n$, on which we introduce a deformation map $\tilde{\vec{\varphi}}:\mathcal{B}_0\times\RVE\rightarrow \mathbb{R}^n$ and the quantities derived from it analogue to the macro continuum. Note, that all quantities on $\Omega_0$ will be marked with a superimposed tilde.  Hence, $\tilde{\vec{\varphi}}$ is defined on a $n+n$ dimensional space.

\subsection{Micro-continuum}
The deformation map $\tilde{\vec{\varphi}}$  is defined as a first order Taylor approximation
\begin{equation}\label{eq:microdefo}
\tilde{\vec{\varphi}}(\vec{X},\tilde{\vec{X}}) = \vec{F}(\vec{X})\tilde{\vec{X}} + \tilde{\vec{w}}(\vec{X},\tilde{\vec{X}}),\quad\vec{X}\in\mathcal{B}_0,\,\tilde{\vec{X}}\in\RVE,
\end{equation} 
such that the micro deformation consists of a homogeneous part $\vec{F}(\vec{X})\tilde{\vec{X}}$ and a non-homogeneous field $\tilde{\vec{w}}\colon\mathcal{B}_0\times\RVE\to\mathbb{R}^n$ referred to as microscopic fluctuations.  Accordingly, the deformation gradient reads
\begin{equation}\label{eq:microGrad}
\tilde{\vec{F}} = \nablaq{\tilde{\vec{X}}}(\tilde{\vec{\varphi}}) = 
\vec{F}(\vec{X}) + \vec{F}',\quad \vec{F}':=\nablaq{\tilde{\vec{X}}}(\tilde{\vec{w}}).
\end{equation}
As usual in multiscale techniques, we relate the macro- and the microscopic deformation gradient and elastic potential via volume averages
\begin{equation}\label{eq:homogenization}
\vec{F}=\frac{1}{|\RVE|}\int\limits_{\RVE}\tilde{\vec{F}}\,\d \tilde{V}\quad\text{and}\quad
\Psi=\frac{1}{|\RVE|}\int\limits_{\RVE}\tilde{\Psi}\,\d \tilde{V},
\end{equation}
where $|\RVE|$ denotes the volume of the RVE. The former requirement is enforced using suitable boundary condition
\begin{equation}\label{eq:boundaryA}
\int\limits_{\partial\RVE}\tilde{\vec{w}}\otimes\tilde{\vec{N}}\,\d \tilde{A} = \vec{0},
\end{equation}
which directly enforces \eqref{eq:homogenization}$_1$. Therefore,  we assume that one of the following conditions is valid
\begin{equation}\label{eq:Con}
\text(i)\;   \tilde{\vec{w}}=\vec{0}\quad\text{in }\RVE, \qquad
\text(ii)\;  \tilde{\vec{w}}=\vec{0}\quad\text{on }\partial\RVE, \qquad 
\text(iii)\; \tilde{\vec{w}}^+=\vec{w}^-\quad\text{on }\partial\RVE,
\end{equation} 
where the first one reconstruct a Voight type homogenization, the second demand homogeneous deformations on the boundary and the last addresses periodic boundary conditions.  Requiring the macro-homogeneity or Hill-Mandel condition 
\begin{equation}\label{eq:hillA}
\frac{1}{|\RVE|}\int\limits_{\RVE}\tilde{\vec{P}}:\delta\tilde{\vec{F}}\,\d \tilde{V} = 
\vec{P}:\delta\vec{F},
\end{equation} 
we can show after a few calculations assuming that the static equilibrium of the micro-continuum is governed by $\nablaq{\tilde{\vec{X}}}\cdot\tilde{\vec{P}} = \vec{0}$ in $\Omega_0$, that this leads to the volume average relationship between macro- and micro-stresses
\begin{equation}\label{eq:averageStress}
\vec{P} = \frac{1}{|\RVE|}\int\limits_{\RVE}\tilde{\vec{P}}\,\d \tilde{V},
\end{equation}
which is equivalent of enforcing \eqref{eq:homogenization}$_2$. 

\subsection{Micro-fracture}
Next, we introduce a phase-field for micro-fracture.  Assuming the absent of macro phase-field $\s:\mathcal{B}_0\rightarrow[0,\,1]$, we introduce a micro phase-field $\ds:\mathcal{B}_0\times\RVE\rightarrow[0,\,1]$, using the space of solution and trial functions
\begin{align}
\mathcal{V}^{\ds} &= \{\ds\in L^2(\mathcal{B}_0;H^1(\RVE))\,|\,\dot{\ds}\geq 0\},\quad
\mathcal{V}^{\delta\ds} = \{\delta\ds\in L^2(\mathcal{B}_0;H^1(\RVE))\},\\
\mathcal{V}^{\tilde{\vec{w}}} &= \{\tilde{\vec{w}}\in L^2(\mathcal{B}_0;H^1(\RVE))\},\quad
\mathcal{V}^{\delta\tilde{\vec{w}}} = \{\delta\tilde{\vec{w}}\in L^2(\mathcal{B}_0;H^1(\RVE))\,|\,\delta\tilde{\vec{w}}=\vec{0}\;\text{on}\;\partial\RVE\}.
\end{align}
This leads to the microscale balance equations
\begin{equation}\label{eq:balanceMicro}
\begin{aligned}
G^{\tilde{\varphi}}:=\int\limits_{\RVE}\frac{\partial\tilde{\Psi}(\tilde{\vec{F}},\ds)}{\partial\tilde{\vec{F}}}:\delta\tilde{\vec{F}}\,\d \tilde{V} & = 0,\\
G^{\ds}:=\int\limits_{\RVE} \delta\ds\left(\frac{\partial\tilde{\Psi}(\tilde{\vec{F}},\ds)}{\partial\ds}+\frac{\tilde{g}_c}{\tilde{l}}\,\ds\right)+ \tilde{g}_c \tilde{l}\,\nabla(\delta\ds)\cdot\nabla(\ds),\d \tilde{V} & = 0,
\end{aligned}
\end{equation}
where $G^{\ds}$ is the total virtual work of the phase-field of the respective $\RVE$.  This allows for pure micro-fracture  and we can calculate the homogenized, macroscopic first Piola-Kirchhoff stress tensor $\vec{P}$ via
\begin{equation}\label{eq:P3}
 \frac{1}{|\RVE|}\int\limits_{\RVE}\frac{\partial\tilde{\Psi}(\tilde{\vec{F}},\ds)}{\partial\tilde{\vec{F}}}\,\d \tilde{V} =: \vec{P},
\end{equation}
analogues to \eqref{eq:P2}. As usual for dissipative and thus, path dependent formulations, we have to store $\ds$ for every single \RVE{}.   

\begin{remark}
Here, we do not apply any kind of Taylor approximation on the macro phase-field, as micro-fracture can be considered as a local dissipation.  More importantly, this kind of microscale fracture can be considered as continuous damage model on the macroscale as presented in, among other, Lemaitre \cite{lemaitre1984,lemaitre1985}. Therein, a scalar damage variable is introduced using the thermodynamic potential $Y = \frac{\partial\Psi}{\partial\s}$ analoguesly to the phase-field driving force $\mathcal{H}$.  The existence of a scalar damage variable is discussed noting that $\s$ \glqq{}represents the surface density of intersection of micro-cracks and micro-cavities with any plane in the body\grqq{}, which are calculated here explicitly as micro-cracks within the RVE. \\
As shown in several publications in the 1990th, the introduction of scalar damage models without consideration of the neighborhood, e.g.\ using an additional gradient term, is posed to instabilities. This is reflected here by the fact, that a full fracture of the RVE using periodic boundary conditions can be considered as infinite large cracks on the microscale, contradicting the assumption of an appropriate scale separation.  As shown in the numerical examples section below, we obtain checkerboard patterns of the phase-field if evaluated on the macroscale within areas of large gradients of the deformation gradient. This has already been described by Peerlings et al. \cite{geers2001}, stating that the \glqq{}numerical analysis [...] becomes sensitive to the discretization\grqq{},  demonstrated here by the checkerboard modes. Thus, we observe that the curvature of the deformation restricts the validity of the chosen approach.
\end{remark}

\subsection{Macro-fracture}
Assuming, that a localization results become physically unrealistic and sensitive to the discretization, nonlocal phase-field approaches on the macroscale become of interest.   Basically,  we assume the existence of the macro phase-field $\s:\mathcal{B}_0\rightarrow[0,\,1]$ and the absence of the micro phase-field $\ds:\mathcal{B}_0\times\RVE\rightarrow[0,\,1]$, i.e.\ we solve 
\begin{equation}\label{eq:P1}
G^{\tilde{\varphi}}:=\int\limits_{\RVE}\frac{\partial\tilde{\Psi}(\tilde{\vec{F}},\s)}{\partial\tilde{\vec{F}}}:\delta\tilde{\vec{F}}\,\d \tilde{V} = 0,  
\end{equation}
where $G^{\tilde{\varphi}}$ is  the total virtual work of the mechanical field on the microscale using a constant macroscale $\s$ within each RVE.  Then, we evaluate the macro stresses again using the volume average
\begin{equation}\label{eq:P2}
 \frac{1}{|\RVE|}\int\limits_{\RVE}\frac{\partial\tilde{\Psi}(\tilde{\vec{F}},\s)}{\partial\tilde{\vec{F}}}\,\d \tilde{V} =: \vec{P},
\end{equation}
For the macro phase-field driving force, we also evaluate the volume average
\begin{equation}\label{eq:H}
\frac{1}{|\RVE|}\int\limits_{\RVE}\tilde{\mathcal{H}}\,\d \tilde{V} =  \mathcal{H},\quad\text{where}\quad\tilde{\mathcal{H}} :=\frac{\partial\tilde{\Psi}(\tilde{\vec{F}},\s)}{\partial\s}.
\end{equation}

\subsection{Dissipation}
Both approaches for pure micro- and pure macro-fracture can be combined in a common framework. Therefore, we introduce a new phase-field variable on the microscale as $\tilde{\s} = \s + \ds$, noting that only one of both, either $\s$ or $\ds$ can be active and the other is set to zero.  Using the degeneration function $g(\tilde{\s}) = 1-(\s + \ds)$ yields
\begin{equation}
\vec{F}^\mathrm{e} = \sum\limits_{a=1}^d(\lambda_a^+)(\lambda_a^+)^{-\s}(\lambda_a^+)^{-\ds}(\lambda_a^-)\vec{\mathfrak{n}}_a\otimes\vec{\mathfrak{N}}_a,
\end{equation}
which demonstrates the general character of the triple split of the tensile stretches.  To obtain the dissipated energy within the RVE, we set $\delta\tilde{\vec{\varphi}} = \dot{\tilde{\vec{\varphi}}}$ and $\delta\tilde{\s} = \dot{\tilde{\s}}$. Thus, we obtain
\begin{equation}
\dot{\tilde{\vec{\varphi}}} = \dot{\vec{\varphi}} + \dot{\vec{F}}\tilde{\vec{X}} + \dot{\tilde{\vec{w}}}, \quad \dot{\tilde{\vec{F}}} = \dot{\vec{F}} + \nablaq{\tilde{\vec{X}}}\dot{\tilde{\vec{w}}}
\quad\text{and}\quad\dot{\tilde{\s}} = \dot{\s} + \dot{\ds}.
\end{equation}
Focusing first on the micro-fracture contributions using $\dot{\tilde{\s}} = \dot{\ds}$ gives
\begin{equation}
\begin{aligned}
-\frac{1}{|\RVE|}\int\limits_{\RVE}\dot{\ds}\underbrace{\frac{\partial\tilde{\Psi}(\tilde{\vec{F}},\tilde{\s}}{\partial\tilde{\s}})}_{=\tilde{\mathcal{H}}}\,\d \tilde{V} &= 
\frac{1}{|\RVE|}\int\limits_{\RVE}\dot{\ds}\,\frac{\tilde{g}_c}{\tilde{l}}\,\tilde{\s}+ \tilde{g}_c \tilde{l}\,\nabla(\dot{\ds})\cdot\nabla(\tilde{\s}),\d \tilde{V} \\ 
&= \frac{1}{|\RVE|}\int\limits_{\RVE}\tilde{g}_c\dot{\tilde{\gamma}}_l\,\d \tilde{V} =: \tilde{D}^{\indi{mi}}.
\end{aligned}
\end{equation}
where $\tilde{D}^{\indi{mi}}$ denotes the volumetric average dissipation density of the micro-system. Next, we introduce the volumetric average of the strain energy as $\Psi = \frac{1}{|\RVE|}\int_{\RVE}\tilde{\Psi}\d \tilde{V}$, where $\tilde{\Psi} := \tilde{\Psi}(\vec{F}, \s, \nablaq{\tilde{\vec{X}}}(\tilde{\vec{w}}),\ds)$ and take the temporal derivative with respect to all four arguments as follows
\begin{equation}
\dot{\Psi} = \frac{1}{|\RVE|}\int\limits_{\RVE}\dot{\tilde{\Psi}}\d \tilde{V} = \vec{P}:\dot{\vec{F}} + \mathcal{H}\dot{\s} + 
\frac{1}{|\RVE|}\int\limits_{\RVE}\tilde{\vec{P}}:\nablaq{\tilde{\vec{X}}}\dot{\tilde{\vec{w}}} + \tilde{\mathcal{H}}\dot{\ds}\,\d \tilde{V},
\end{equation}
Insertion into the balance equation on the macroscale in \eqref{eq:conti2} along with \eqref{eq:contiPF} allows us to write for the energy balance
\begin{equation}
\frac{\d}{\d t}T +  \underbrace{\int\limits_{\mathcal{B}_0}\dot{\Psi} + \tilde{D}^{\indi{mi}}\,\d V}_{=P^{\indi{int}} + D^{\indi{mi}}}  + D^{\indi{ma}}  = P^{\indi{ext}}
\end{equation}
Thus, we obtain
\begin{equation}
\frac{\d}{\d t}T +  P^{\indi{int}} + D^{\indi{mi}}  + D^{\indi{ma}}  = P^{\indi{ext}},
\end{equation}
demonstrating the conservation of energy with dissipation on the micro- and the macro\-scale. Hence, pure macro-fracture leads to $D^{\indi{mi}} = 0$, whereas pure micro-fracture sets the macro dissipation $D^{\indi{ma}} = 0$.  Allowing for only one phase-field variable at the same position, i.e.\ either $\s$ or $\ds$ prevents the transfer of energy between both dissipation potentials $D^{\indi{mi}}$ and $D^{\indi{ma}}$, which would result in an unphysical behavior.  Note, that the additive construction of $\tilde{\s} = \s + \ds$ yields the assumed additive construction of the micro and macro dissipation. The switch between both models depends on the scale separation and the gradient of the deformation gradient.  We will demonstrate this in the Section on numerical examples.

%% file: sections/discretization.tex
\section{Discretization and solution strategy}\label{sec:disc}
To achieve a numerical solution for the problem, we apply a finite element framework on both scales. In particular, we consider a standard displacement-based finite element approach, where we introduce finite dimensional approximations of $\vec{\varphi}$ and $\delta\vec{\varphi}$ as well as of $\s$ and $\delta\s$, so that\footnote{Whenever convenient, we make use of the Einstein summation convention on repeated indices.}
\begin{align}
\vec{\varphi}^{\indi{h}} = \sum\limits_{A\in\omega}N^A(\vec{X})\vec{q}_A\quad\text{and}\quad\delta\vec{\varphi}^{\indi{h}} = \sum\limits_{A\in\omega}N^A(\vec{X})\delta\vec{q}_A,\\
\s^{\indi{h}} = \sum\limits_{A\in\omega}N^A(\vec{X})\s_A\quad\text{and}\quad\delta\s^{\indi{h}} = \sum\limits_{A\in\omega}N^A(\vec{X})\delta\s_A.
\end{align}
Here, $A\in\omega=\{1,\,\hdots,\,n_{\text{node}}\}$, such that $\vec{q}_A\in\mathbb{R}^n$ denotes the position vector of node $A$ and $N^A(\vec{X}):\mathcal{B}_0\rightarrow\mathbb{R}$ are global shape functions. Within each RVE, we introduce the finite dimensional approximations of the fluctuations and the local phase-field as
\begin{align}
\tilde{\vec{w}}^{\indi{h}} = \sum\limits_{s\in\tilde{\omega}}R^s(\tilde{\vec{X}})\tilde{\vec{q}}_s\quad\text{and}\quad\delta\tilde{\vec{w}}^{\indi{h}} = \sum\limits_{t\in\tilde{\omega}}R^t(\tilde{\vec{X}})\delta\tilde{\vec{q}}_t,\\
\ds^{\indi{h}} = \sum\limits_{s\in\tilde{\omega}}R^s(\tilde{\vec{X}})\ds_s\quad\text{and}\quad\delta\ds^{\indi{h}} = \sum\limits_{t\in\tilde{\omega}}R^t(\tilde{\vec{X}})\delta\ds_t.
\end{align}

As usual for multiscale fracture, we apply a staggered formulation of the Newton-Raphson iteration: First,  we solve exclusively for $\tilde{\vec{w}}^{\indi{h}}$ and $\ds^{\indi{h}}$. Hence,  we solve within each iteration step until either \eqref{eq:balanceMicro} or \eqref{eq:P1} is fulfilled. Therefore, we can correlate all $\Delta\tilde{\vec{q}}_t$ and $\Delta\ds_t$, $t\in\tilde{\omega}$ within each RVE to the macrovalues $\Delta\vec{F}$ and $\Delta\s$, evaluated at the corresponding Gauss node,  and obtain afterwards 
\begin{equation}\label{eq:delta_w}
\int\limits_{\RVE} \nablaq{\tilde{X}}(\delta\tilde{\vec{w}}^{\indi{h}}):\mathbb{C}_{\mathrm{h}}:\left[\Delta \vec{F} + \nablaq{\tilde{\vec{X}}}(\Delta\tilde{\vec{w}}^{\indi{h}})\right] + 
\nablaq{\tilde{X}}(\delta\tilde{\vec{w}}^{\indi{h}}):\mathbb{D}_{\mathrm{h}}\left[\Delta \s + \Delta \tilde{d}^{\indi{h}}\right]\d V = 0,
\end{equation}
\begin{equation}
\begin{aligned}
\int\limits_{\RVE} \delta\tilde{d}^{\indi{h}}\,\mathbb{D}^T_{\mathrm{h}}:\left[\Delta \vec{F} + \nablaq{\tilde{\vec{X}}}(\Delta\tilde{\vec{w}}^{\indi{h}})\right] + 
\delta\tilde{d}^{\indi{h}}\,\mathbb{Q}_{\mathrm{h}}\left[\Delta \s + \Delta \tilde{d}^{\indi{h}}\right] + &\\
\delta\tilde{d}^{\indi{h}}\frac{\tilde{g}_c}{\tilde{l}}\Delta\tilde{d}^{\indi{h}} + \tilde{g}_c \tilde{l}\, \nabla(\delta\tilde{d}^{\indi{h}})\cdot&\nabla(\Delta\tilde{d}^{\indi{h}})\d V = 0,\label{eq:delta_s}
\end{aligned}
\end{equation}
noting that $\nabla(\Delta\s) = \vec{0}$ within each RVE and
\begin{equation}
\mathbb{C}_{\mathrm{h}} := \frac{\partial\tilde{\vec{P}}}{\partial\tilde{\vec{F}}},\quad\mathbb{D}_{\mathrm{h}} := \frac{\partial\tilde{\vec{P}}}{\partial\tilde{\s}}\quad\text{and}\quad 
\mathbb{Q}_{\mathrm{h}} := \frac{\partial\tilde{\mathcal{H}}}{\partial\tilde{\s}}.
\end{equation}
Thus, we obtain
\begin{equation}
\begin{bmatrix}
\delta \tilde{\vec{q}}_s & \delta \ds_s
\end{bmatrix}
\vec{\mathcal{M}}^{st}
\begin{bmatrix}
\Delta\tilde{\vec{q}}_t \\
\Delta \ds_t
\end{bmatrix} = -
\begin{bmatrix}
\delta \tilde{\vec{q}}_s & \delta \ds_s
\end{bmatrix}
\vec{\mathcal{K}}^s
\begin{bmatrix}
: \Delta\vec{F} \\
\Delta\s
\end{bmatrix},
\end{equation}
where
\begin{equation}
\vec{\mathcal{M}}^{st} = 
\begin{bmatrix}
\int\limits_{\RVE}\nabla\tilde{{R}}^{s}\cdot\mathbb{C}_{\mathrm{h}}\nabla\tilde{{R}}^{t}\d V & 
\int\limits_{\RVE}\nabla\tilde{{R}}^{s}\cdot\mathbb{D}_{\mathrm{h}}\tilde{R}^{t}\d V\\
\int\limits_{\RVE}\tilde{R}^{s}\,\mathbb{D}^T_{\mathrm{h}}\,\nabla\tilde{{R}}^{t}\d V & 
\int\limits_{\RVE}\tilde{R}^{s}\,\mathbb{Q}_{\mathrm{h}}\tilde{R}^{t} + \tilde{R}^{s}\,\frac{\tilde{g}_c}{\tilde{l}}\tilde{R}^{t} + \tilde{g}_c\tilde{l}\,\nabla\tilde{R}^{s}\cdot\nabla\tilde{R}^{t}\d V
\end{bmatrix},
\end{equation}
and
\begin{equation}
\vec{\mathcal{K}}^s = 
\begin{bmatrix}
\int\limits_{\RVE}\nabla\tilde{{R}}^{s}\cdot\mathbb{C}_{\mathrm{h}}\d V & 
\int\limits_{\RVE}\nabla\tilde{{R}}^{s}\cdot\mathbb{D}_{\mathrm{h}}\d V\\
\int\limits_{\RVE}\tilde{R}^{s}\,\mathbb{D}^T_{\mathrm{h}}\d V & 
\int\limits_{\RVE}\tilde{R}^{s}\,\mathbb{Q}_{\mathrm{h}}\d V
\end{bmatrix}.
\end{equation}
This gives us the sensitivity
\begin{equation}\label{eq:transfer}
\begin{bmatrix}
\Delta\tilde{\vec{q}}_t \\
\Delta \ds_t
\end{bmatrix} = -
\left(\vec{\mathcal{M}}^{-1}\right)^{ts}\vec{\mathcal{K}}^s
\begin{bmatrix}
: \Delta\vec{F} \\
\Delta\s
\end{bmatrix}.
\end{equation}
This can now be inserted into the remaining macroscale balance equations to derive a condensed tangent matrix for the solution of the macro-values, which can be considered as null-space projection as shown in \cite{hesch2024}. With this, the next step of the staggered solution can be started. Note that a monolithic solution is presented in \cite{hesch2024},  which avoids the staggered scheme. 

\begin{remark}
The solution on both scales can be replaced by an additional staggered solution between mechanical field and phase-field. First, the mechanical field and the phase field can be solved successively on the micro-scale, until  \eqref{eq:balanceMicro}  is valid, which allows to make use of \eqref{eq:transfer}.  The macro-scale can than be solved monolithically. Second, a staggered micro-macro formulation can be applied, by solving the mechanical field and the phase-field successively on both scales, i.e.\ we solve  \eqref{eq:balanceMicro}$_1$, transfer the mechanical information on the macro-scale using \eqref{eq:delta_w} and solve the macro mechanical field. Afterwards, we solve  \eqref{eq:balanceMicro}$_2$, and make use of \eqref{eq:delta_s} for the transfer. Hence, the off diagonal matrices in \(\left(\vec{\mathcal{M}}^{st}\right)^{-1}\vec{\mathcal{K}}^s\) are erased from the transfer operator in \eqref{eq:transfer}.

\end{remark}

%% file: sections/numerics.tex
\section{Numerical examples}\label{sec:numerics}
In this Section, we investigate the proposed formulations for multiscale fracture problems. The micro-scale constitutive behavior is assumed to be governed by a compressible Mooney-Rivlin material. Without loss of generality, we consider the special case of a two-dimensional plane strain assumption to reduce the computational load. Therefore, we introduce the associated three-dimensional strain energy density function given by
\begin{equation}
\tilde{\Psi} = \alpha(I_1 -3) + \beta (I_2 -3) -2(\alpha+2\beta)\op{ln}(J) + \frac{\lambda}{2}(J-1)^2,
\end{equation}
where $J = \op{det}(\vec{F})$, $I_1 = \vec{F}:\vec{F}$ and $I_2 = \vec{H}:\vec{H}$,  $\vec{H} := \op{cof}(\vec{F})$.  In the two-dimensional case we assume
\begin{equation}
\nabla_{\vec{X}}\vec{\varphi} = 
\begin{bmatrix}
\vec{F}_{2D} & \vec{0} \\ \vec{0} & 1
\end{bmatrix},
\end{equation}
for the plane-strain condition. Hence, we obtain (cf.  Hesch et al. ???hesch2017a for details)
\begin{equation}
\tilde{\Psi} = (\alpha + \beta)(\vec{F}_{2D}:\vec{F}_{2D} -2) + \beta(j^2-1) - (\alpha+2\beta)\op{ln}(j) + \frac{\lambda}{2}(j-1)^2,
\end{equation}
using $j = \op{det}(\vec{F}_{2D})$.

\begin{figure}[htb]
\begin{center} 
	\psfrag{w}[tc][bc]{\footnotesize \(120/128\;[mm]\)}
	\psfrag{b}[c][c]{\footnotesize \(760/128\;[mm]\)}
	\psfrag{l}[c][c]{\footnotesize \(5\;[mm]\)} %\vspace{2cm}
\includegraphics[width=0.5\textwidth]{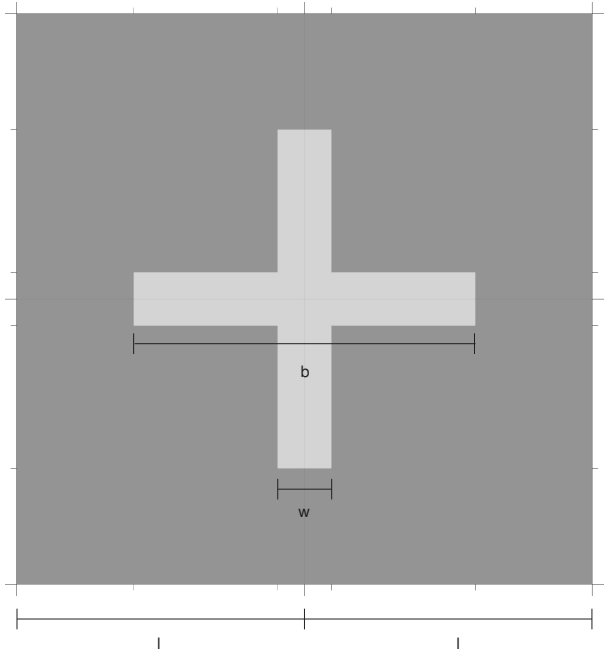} 
\caption{Geometry of the axisymmetric RVE.}\label{fig:RVE_geometry}
\end{center}
\end{figure}

The RVE consists of two different materials with perfect contact interfaces: The material constants of the surrounding matrix material are assumed to take the values $c_1 = 25.64\times 10^3\;[\op{N}/\op{mm}^2]$,  $c_2 = 12.82\times 10^3\;[\op{N}/\op{mm}^2]$ and $c = 57.69\times 10^3\;[\op{N}/\op{mm}^2]$, whereas the inclusion is governed by $c_1 = 12.82\times 10^3\;[\op{N}/\op{mm}^2]$,  $c_2 = 6.41\times 10^3\;[\op{N}/\op{mm}^2]$ and $c = 28.85\times 10^3\;[\op{N}/\op{mm}^2]$, see Figure \ref{fig:RVE_geometry} for details on the geometrical layout.

\subsection{Shear test, micro-fracture}\label{sec:shear_micro}

\begin{figure}[htb]
\begin{center} 
\psfrag{5}[][]{\footnotesize{\begin{footnotesize}\(500\;[mm]\)\end{footnotesize}}}
\psfrag{u}{\footnotesize{\(\vec{u}\)}}
\includegraphics[width=0.5\textwidth]{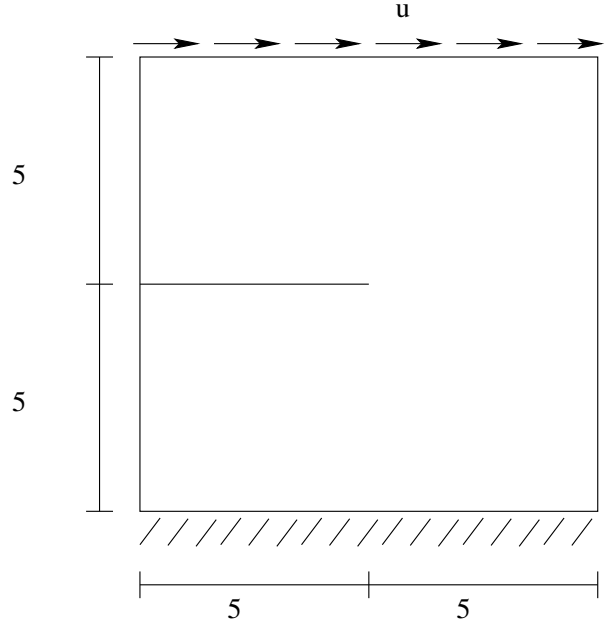} 
\caption{Geometry and boundary conditions for the shear problem.}\label{fig:shear}
\end{center}
\end{figure}

A well-established benchmark problem is that of a squared plate with a horizontal notch. In particular, we make use of duplicated, non-connected nodes along the initial, horizontal notch. The geometry and boundary conditions of the shear test example are depicted in Figure \ref{fig:shear}. The lower boundary is completely constrained, such that no displacement is allowed. The upper boundary is incrementally, uniformly displaced in horizontal direction using an increment of $\Delta\vec{u} = 1\times 10^{-1}\op{mm}$.

The macro-phase field $\s$ is not taken into account.  For the micro-fracture we apply a specific fracture energy of $g_c = 2.5\times 10^4\;[\op{N}/\op{mm}]$ for the matrix material and $g_c = 1.25\times 10^4\;[\op{N}/\op{mm}]$ for the inclusion. The length scale $l = 0.15625 [mm]$ is adjusted for both, the matrix and the inclusion, to the same length such that $l \approx 2h$. The material constants of the surrounding matrix material is the same as in the previous example, whereas the inclusion is governed by $c_1 = 2.56\times 10^3\;[\op{N}/\op{mm}^2]$,  $c_2 = 1.28\times 10^3\;[\op{N}/\op{mm}^2]$ and $c = 5.77\times 10^3\;[\op{N}/\op{mm}^2]$.  The parameters are set in such a way, that we expect the fracture initializes within the inclusion. 

\begin{figure}[H]
\begin{center} 
\includegraphics[width=0.485\textwidth]{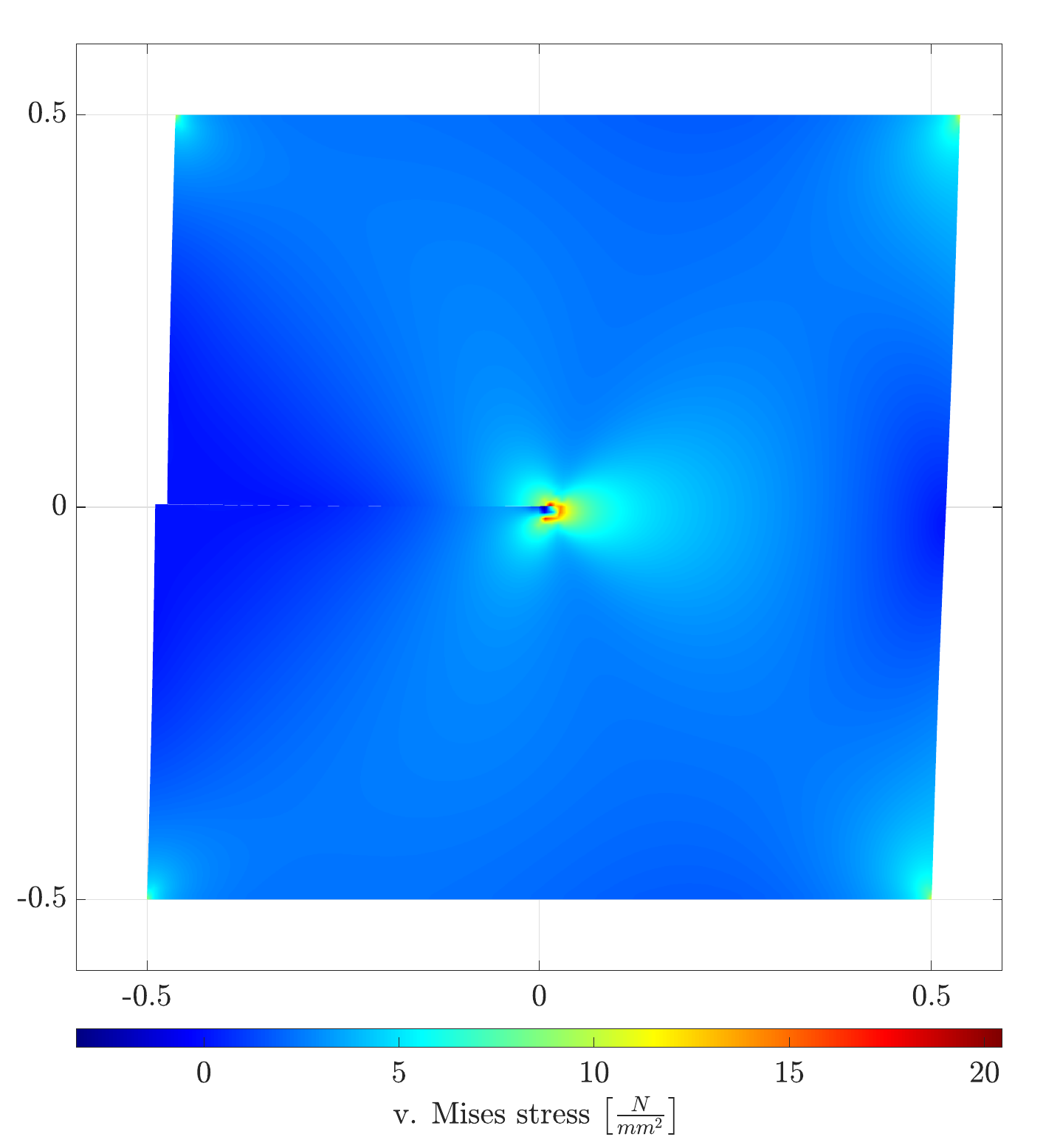} 
\caption{Micro-fracture: Homogenized von Mises stresses at $\Delta\vec{u} = 36.6\op{mm}$.}\label{fig:microF_macro_stress}
\end{center}
\end{figure}

\begin{figure}[H]
\begin{center} 
\includegraphics[width=0.485\textwidth]{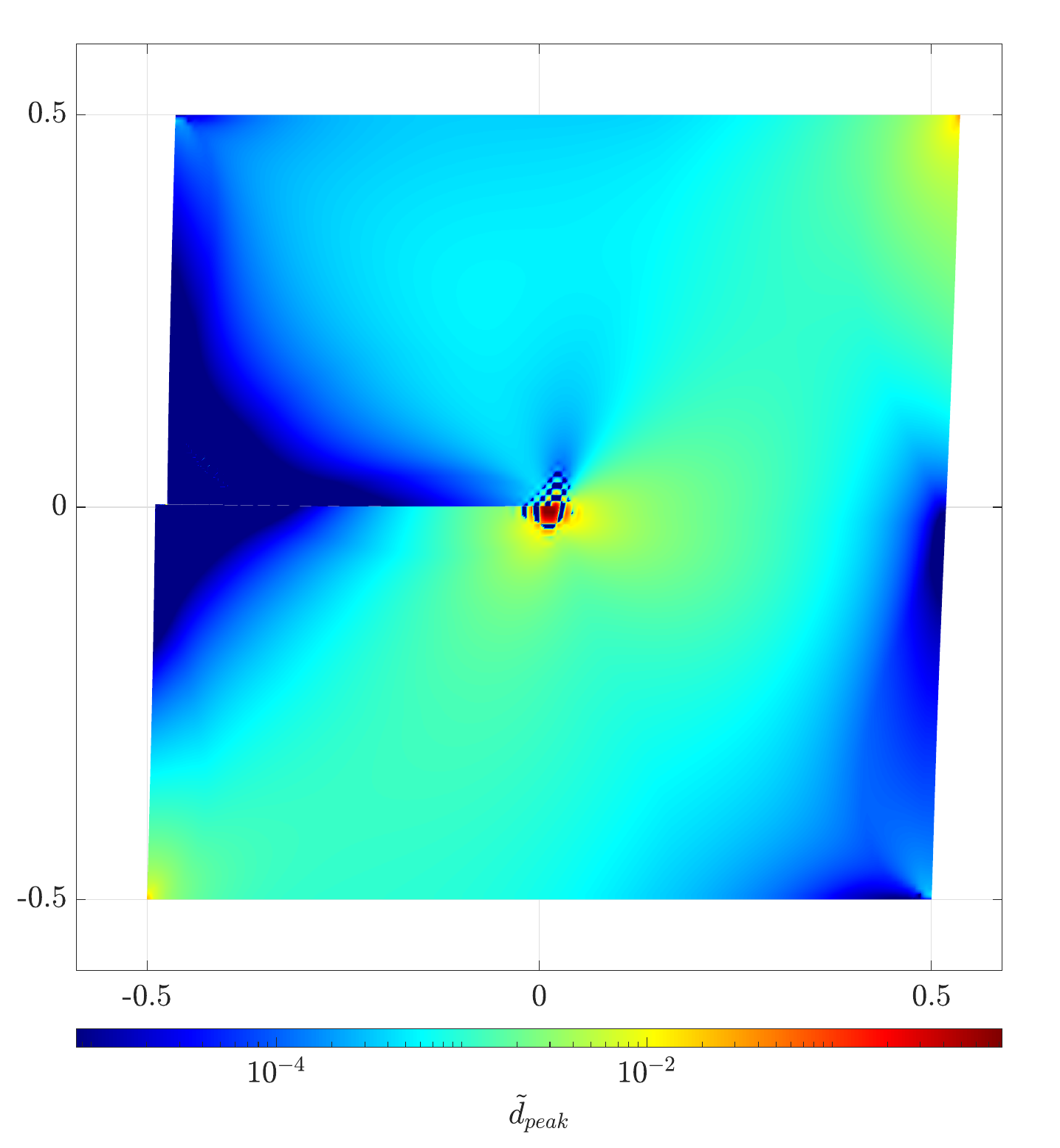} 
\caption{Micro-fracture: Peak values of the micro phase-field for each RVE/GP and interpolated for the macro-scale, plotted in the actual configuration at $\Delta\vec{u} = 36.6\op{mm}$.}\label{fig:microF_macro_peakPhase}
\end{center}
\end{figure}

In figure  \ref{fig:microF_macro_stress}, left the stress distributions at $\Delta\vec{u} = 36.6\op{mm}$ is displayed, whereas in figure \ref{fig:microF_macro_peakPhase} the peak values of the RVE’s individual local phase-fields are shown. As can be seen, we obtain a wide area of damage without the formation of a regularized fracture interface at the macro-scale.

\begin{figure}[H]
\begin{center}  
\includegraphics[width=0.485\textwidth]{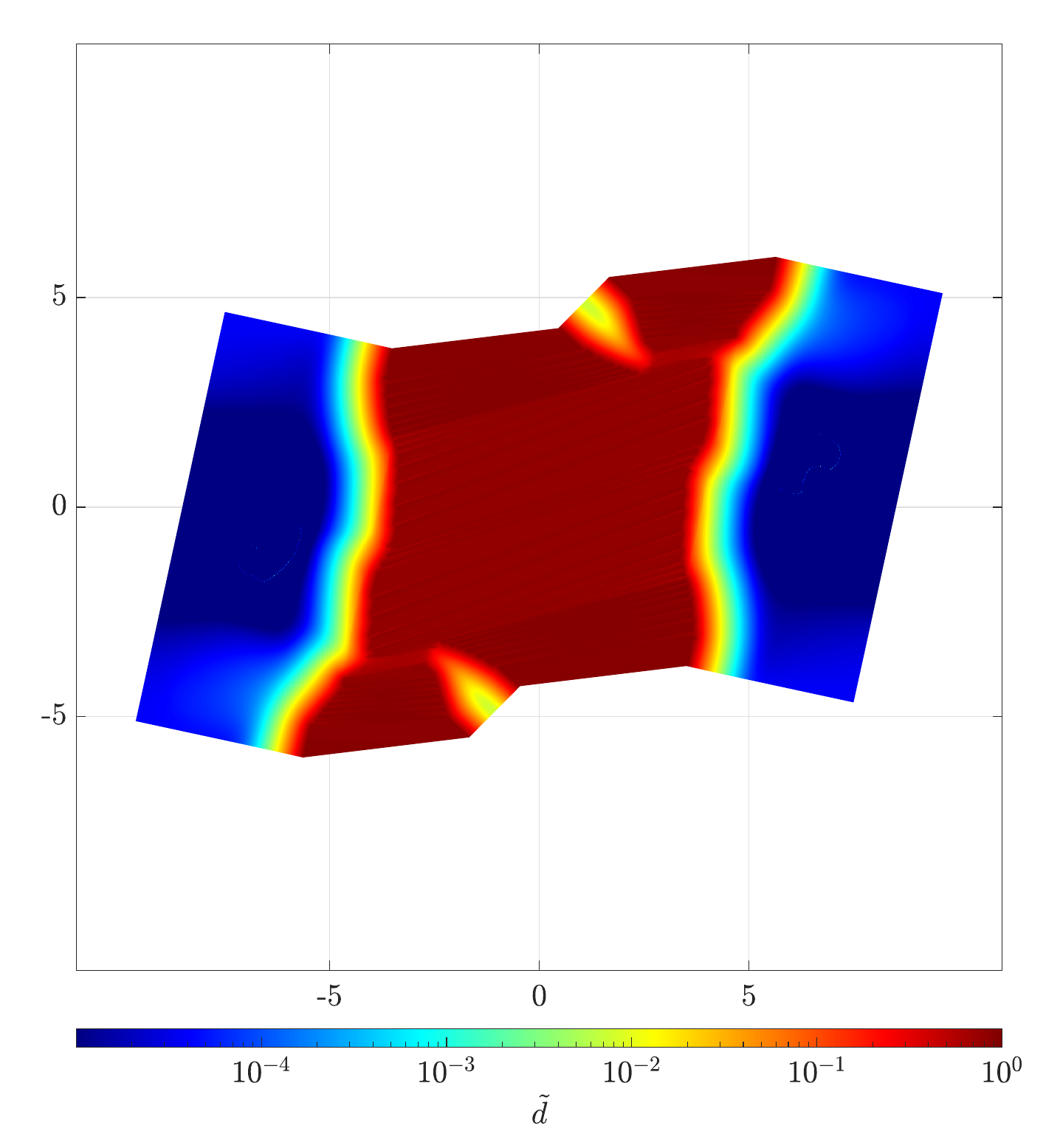} 
\includegraphics[width=0.485\textwidth]{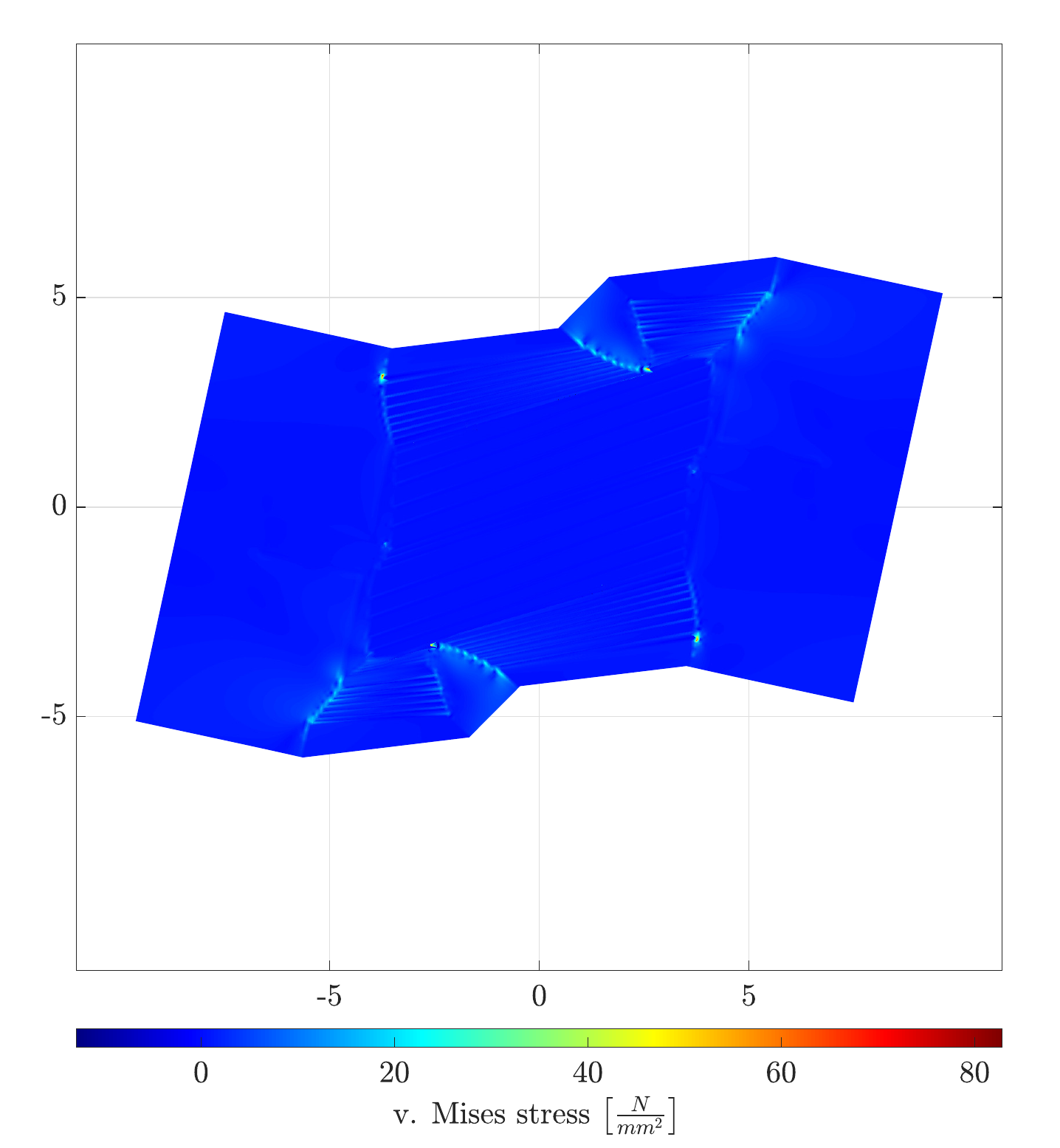} 
\caption{Micro-fracture: Phase-field (left) and von Mises stresses (right) for RVE 32512 at $\Delta\vec{u} = 36.6\op{mm}$.}\label{fig:microF_RVE_phaseStress}
\end{center}
\end{figure}

Moreover, we observe an instability in the direct area around the tip of the horizontal notch showing checkerboard modes in the phase-field. This is in line with the predictions in Peerings et al.  \cite{geers2001} and other. Interestingly, the von Mises stress distribution in figure  \ref{fig:microF_macro_stress}, left, is homogeneous. Searching again for the RVE with the maximum deformation, we obtain in Figure \ref{fig:microF_RVE_phaseStress}  left, the local phase-field and right, the local von Mises stress distribution.

Obviously, the inclusion is fractured in one direction, i.e. we obtain an anisotropic stress degeneration in tension orthogonal to the fracture. Note, that beside the inclusion the surrounding material also fractured and we obtain in the respective direction a complete degeneration of the stresses. Using periodic boundary conditions, this crack extends infinitely in vertical direction, i.e. the assumption of a suitable scale separation does not hold any more, resulting in the instability shown on the macroscale. Hence, we observe graphically the limit of a local damage model without gradient contributions.

\subsection{Cook's membrane, micro-fracture}
\begin{figure}[ht]
\begin{center}
\footnotesize
\begin{tabular}{cc}
\includegraphics[width=0.3825\textwidth]{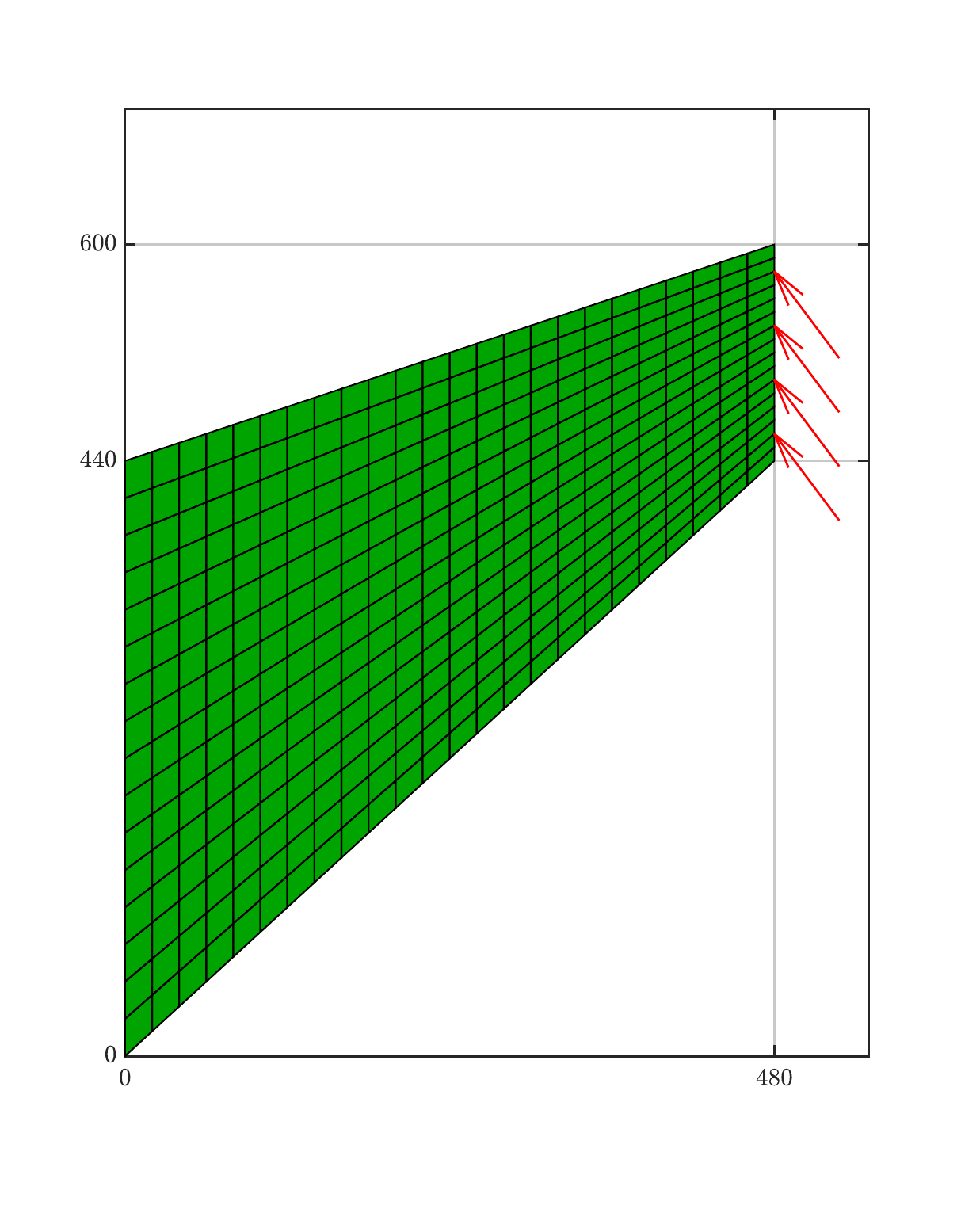}
\end{tabular}
\end{center}
\caption{Reference configuration and computational mesh of the macro-system. The 
red arrows indicate the impact of the surface load defined in Eq.\ \eqref{eq:load}. }
\label{fig:refCon}
\end{figure}

In the previous benchmark example,  the predefined horizontal notch introduces a singular point on which every mesh refinement produces higher stresses.  In this example we demonstrate the applicability on the well-known Cook's membrane.  The reference configuration of the macro-continuum is defined by the four 
points $\vec{P}_1=(0,0)^\mathrm{T}$, $\vec{P}_2=(480,440)^\mathrm{T}$, $\vec{P}_3=(480,600)^\mathrm{T}$, 
$\vec{P}_4=(0,440)^\mathrm{T}$, where $\vec{P}_1$ denotes the lower left, $\vec{P}_2$ the lower right, 
$\vec{P}_3$ the upper right and $\vec{P}_4$ the upper left corner of the membrane. Consequently, the computational  
macro-domain is given by
\begin{equation}\label{eq:refcon}
\mathcal{B}_0 = \left\{\vec{X}=(X_1,X_2)\in\mathbb{R}^2\, |\, 
X_1\in(0,480),\, \frac{11}{12}\,X_1<X_2<\frac{1}{3}\,X_1+440 \right\}
\end{equation}
Regarding the boundary conditions, we assume that the left edge 
$\Gamma^{\varphi}=\{\vec{X}\in\mathbb{R}^2\, | \, X_1=0,\, 0<X_2<440\}$ of the membrane is fixed, 
whereas the remaining boundary $\Gamma^{\sigma}=\partial\mathcal{B}_0\backslash \Gamma^{\varphi}$ 
is exposed to the surface load 
\begin{equation}\label{eq:load}
\vec{T}\colon\ \Gamma^{\sigma}\to\mathbb{R}^2,\quad
\vec{T}(\vec{X}) = \begin{cases}
(-30,40)^\mathrm{T} & \text{if } \vec{X}\in\Gamma^{\sigma}_r, \\
(0,0)^\mathrm{T}   & \text{otherwise},
\end{cases}
\end{equation}
acting on the right edge 
$\Gamma^{\sigma}_r := \{\vec{X}\in\mathbb{R}^2\, | \, X_1=480,\, 440<X_2<600\}$ of $\mathcal{B}_0$, 
see Figure~\ref{fig:refCon} for illustration. For the load application, we use an incremental scheme so that in each case the load is increased in the form $\vec{T}_k:=k\vec{T}$ for the $k$th step.
%%%
%%% Reference configuration
%%%
%%%
Moreover, we assume a zero body load ($\vec{B}=\vec{0}$) and we use the same RVE as in Section \ref{sec:shear_micro}.

\begin{figure}[th]
\begin{center}  
\includegraphics[width=0.485\textwidth]{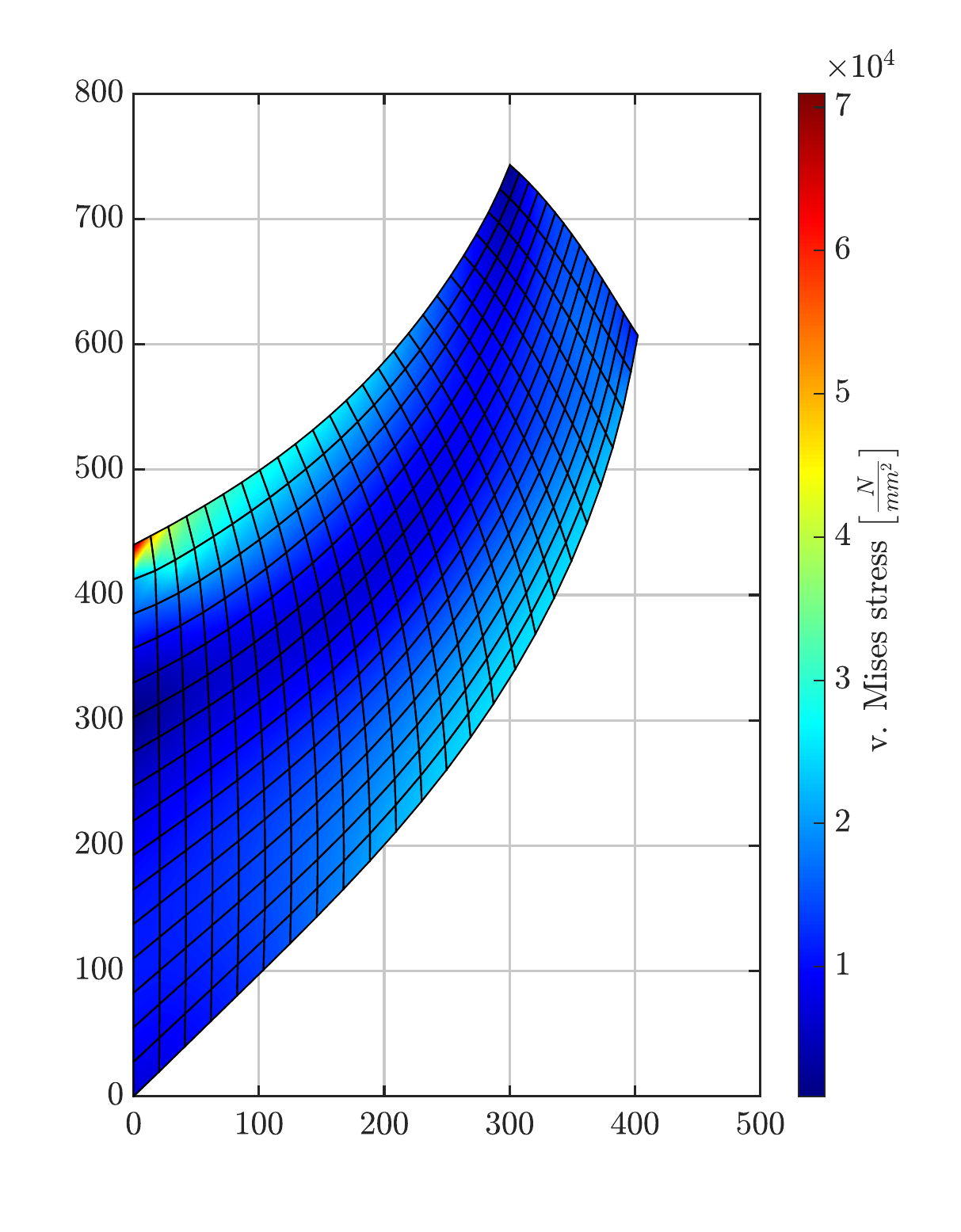} 
\includegraphics[width=0.485\textwidth]{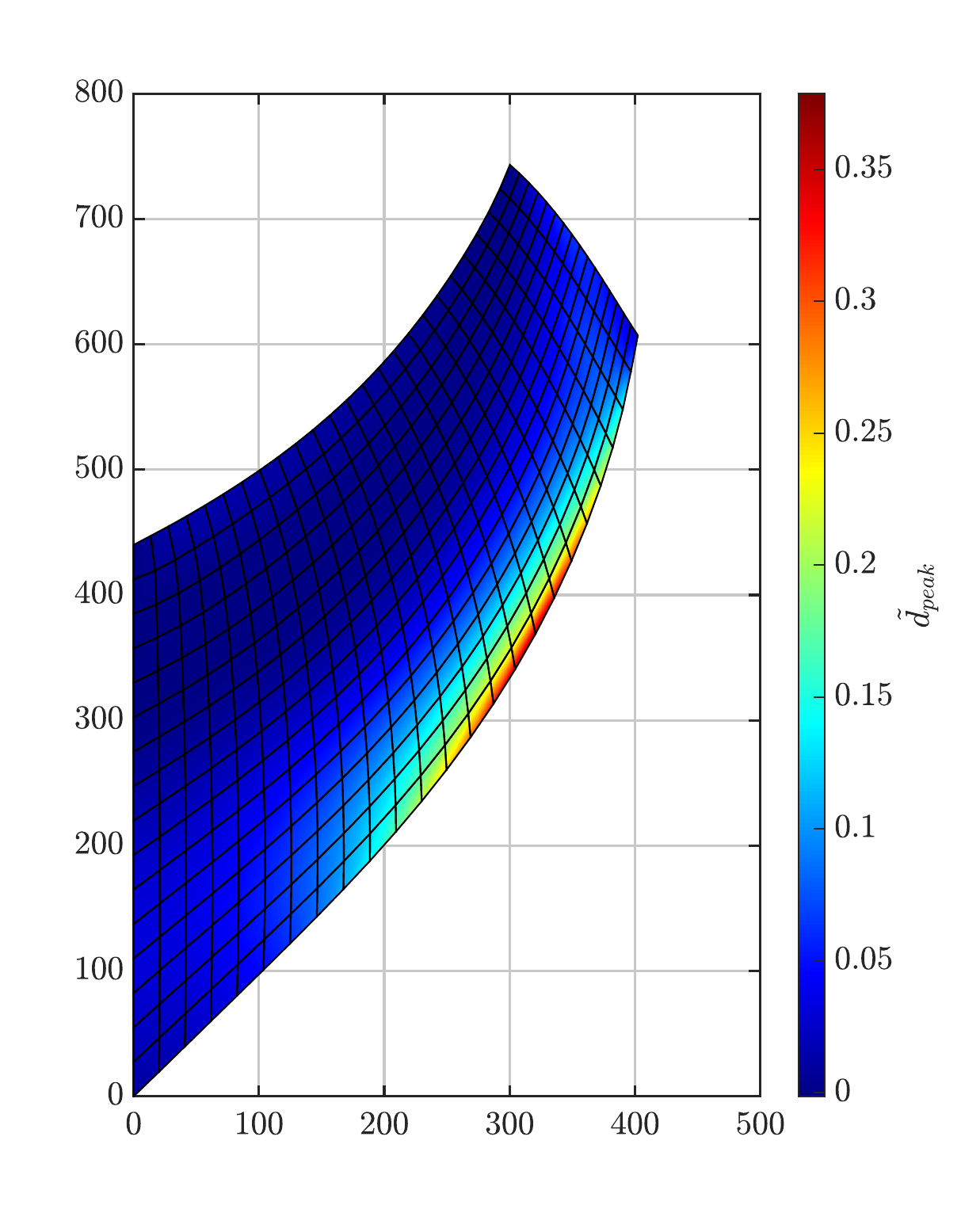} 
\caption{Micro-fracture: Deformed Cook's membrane  at 128 load increments. Left: Homogenized von Mises stresses. Right: Peak values of the micro phase-field.}\label{fig:micro_RVE_stress2}
\end{center}
\end{figure}

As shown in Figure \ref{fig:micro_RVE_stress2}, left, we obtain a stress concentration on the upper left point. However the phase-field will not react on a state of compression, i.e.\ we expect damage only on the lower side of the membrane as can be observed in Figure \ref{fig:micro_RVE_stress2}, right. Here, no instabilities are observed as the peak phase-field is below 0.4. Moreover, no large gradients of the deformation gradient in the state of tension exist.

\subsection{Shear test, macro-fracture}
The geometry and boundary conditions of the macro-fracture shear test example are the same as for the previous micro-fracture test. The macro-scale is discretized using $128^2$ linear elements with a four point Gauss integration.  The specific fracture energy of the macro-scale is given by $g_c = 2.5\times 10^4\;[\op{N}/\op{mm}]$. The length scale $l = 1.5625 [mm]$ is adjusted such that $l \approx 2h$, where h is a characteristic  element length. 

\begin{figure}[H]
\begin{center}  
\includegraphics[width=0.485\textwidth]{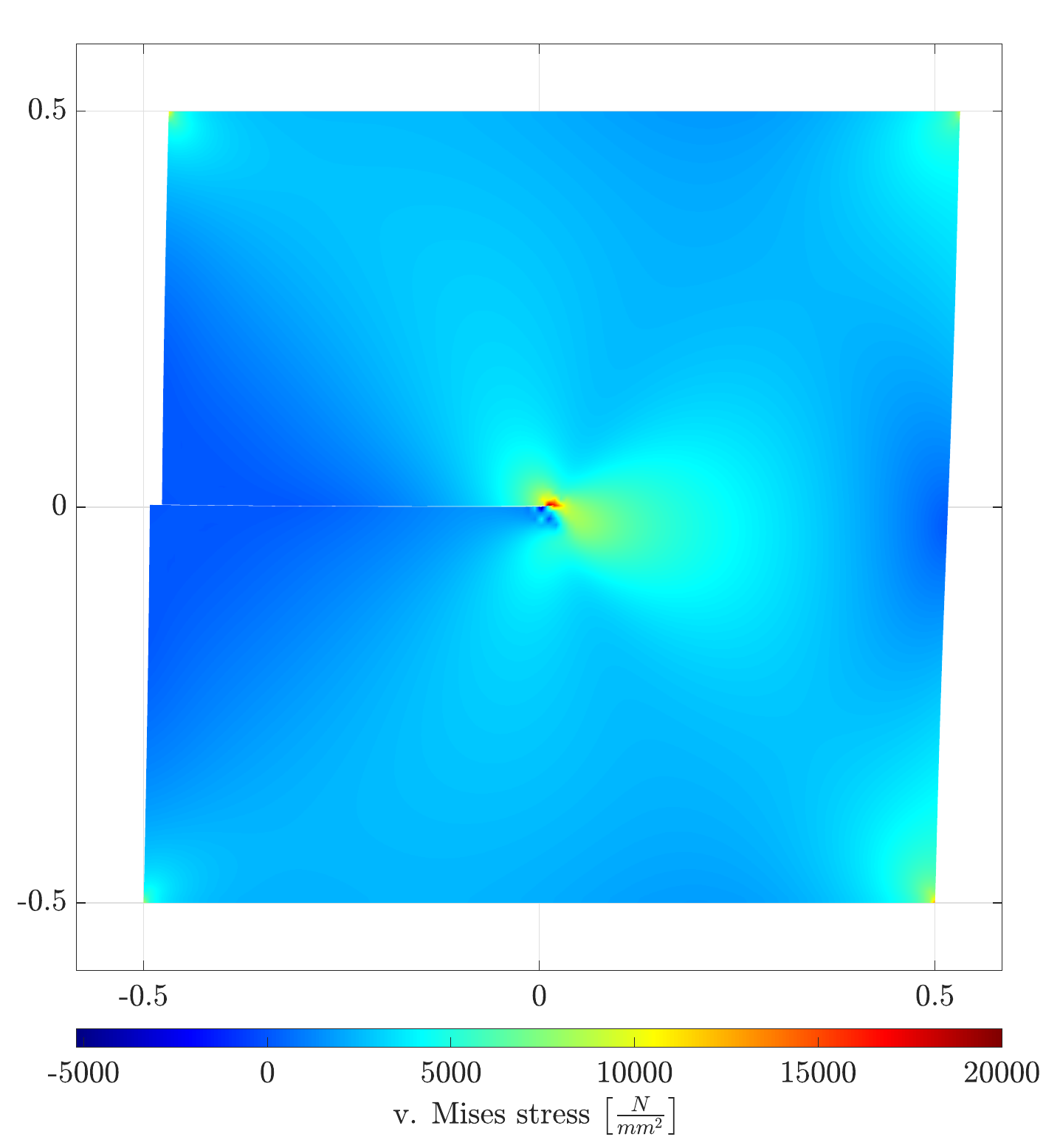} 
\caption{Macro-fracture: Homogenized von Mises stresses at $\Delta\vec{u} = 32\op{mm}$.}\label{fig:macroF_macro_stressDrvngFrc}
\end{center}
\end{figure}

\begin{figure}[H]
\begin{center} 
\includegraphics[width=0.485\textwidth]{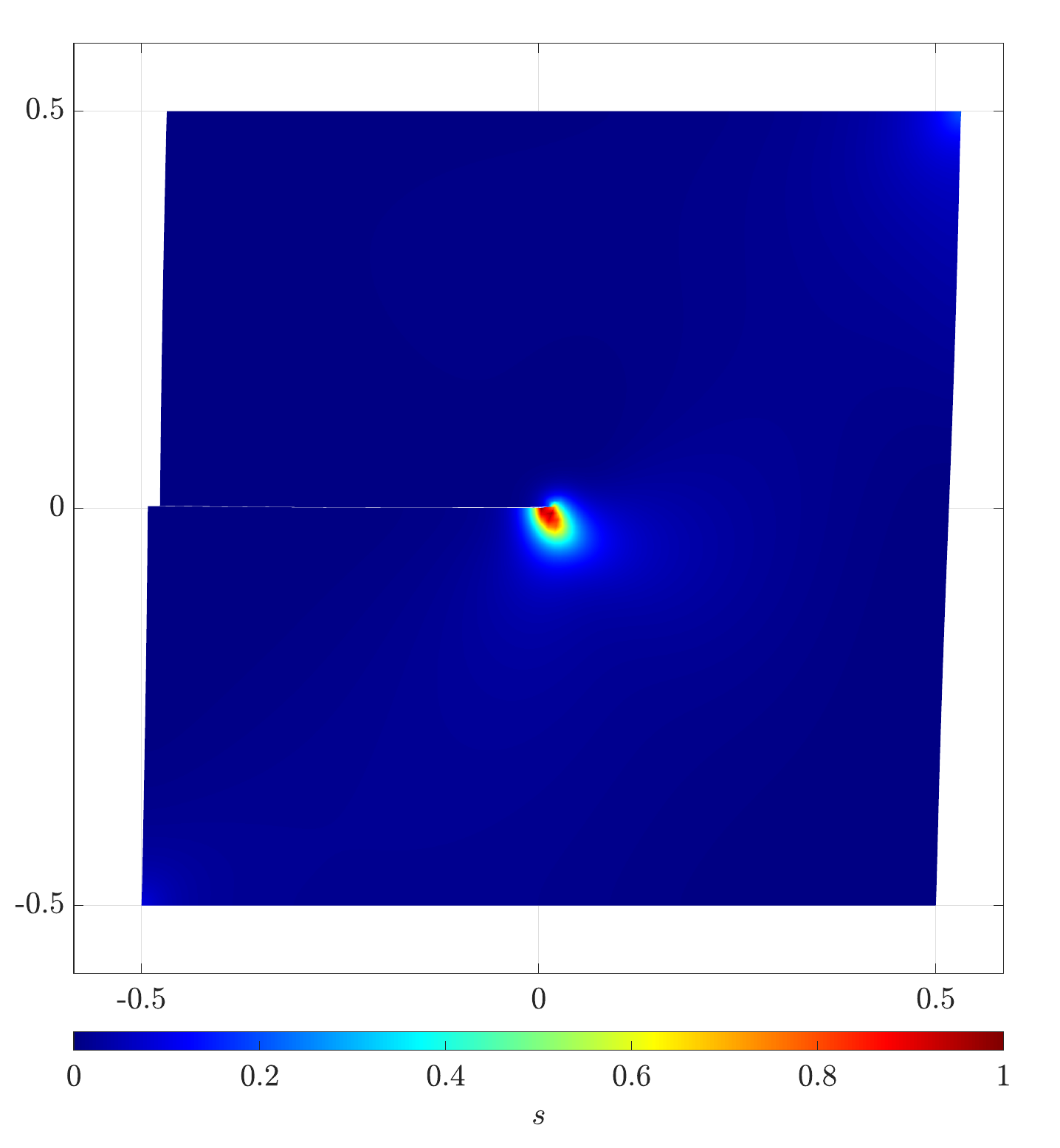} 
\caption{Macro-fracture: Phase-field at the macro-scale, plotted in the actual configuration at $\Delta\vec{u} = 32\op{mm}$.}\label{fig:macroF_macro_phase}
\end{center}
\end{figure}

The von Mises stresses and the phase-field driving force, i.e. both volume averages of the RVEs are shown in Figure \ref{fig:macroF_macro_stressDrvngFrc} for $\Delta\vec{u} = 31\op{mm}$. Moreover, the resulting phase-field is presented in figure \ref{fig:macroF_macro_phase}. Fracture initializes in the position and direction as we expect, cf.\ \cite{hesch2014} where we make use of the same example using a phenomenological strain energy function on a single scale.

\begin{figure}[th]
\begin{center}  
\includegraphics[width=0.485\textwidth]{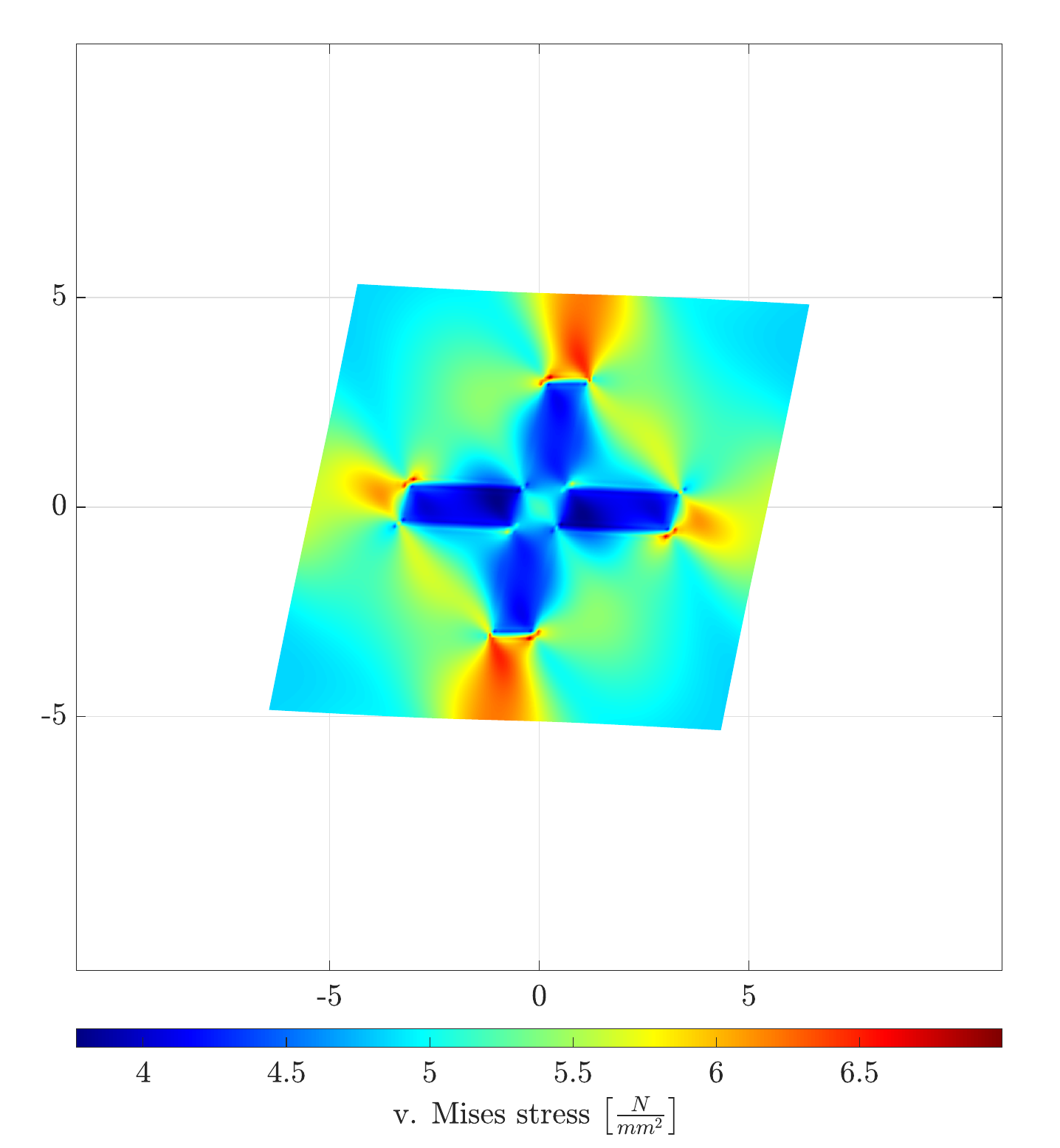} 
\includegraphics[width=0.485\textwidth]{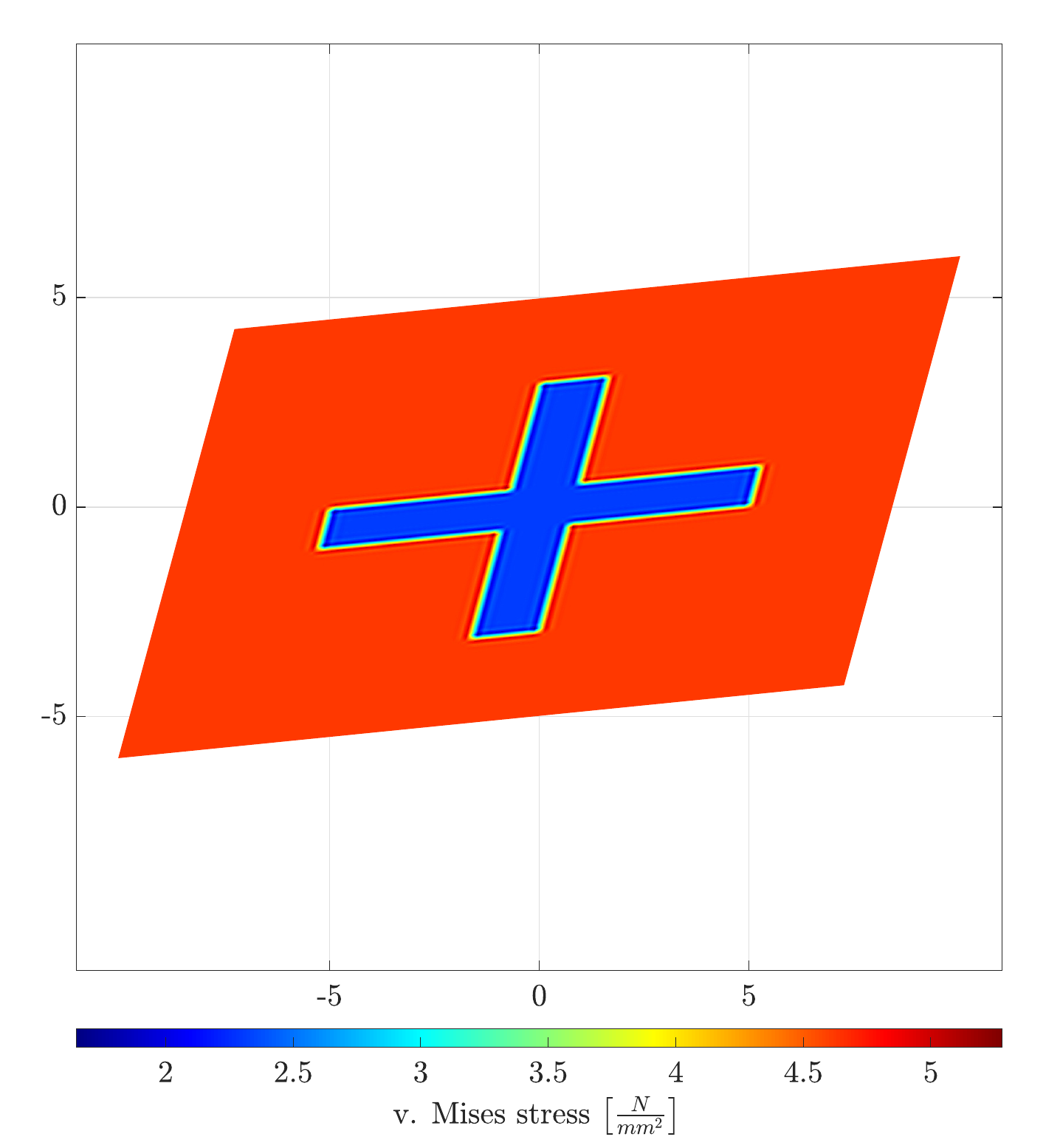} 
\caption{Macro-fracture: Von Mises stress distribution within the deformed RVE at $\Delta\vec{u} = 31\op{mm}$. Left: RVE searching for the maximum deformation and phase-field below threshold, right: A different RVE searching for the maximum deformation above the threshold.}\label{fig:macro_RVE_stress}
\end{center}
\end{figure}

Note, that the local deformation within the fracture zone can become extremely large as the local stresses are reduced with regard to high values of the phase-field, effecting the convergence of the Newton-Raphson iteration to solve the balance equation within the RVE. In particular, local fluctuations within an RVE subject to high macro phase-field values $\s$ become meaningless, as micro-deformation pattern within a (nearly) broken area cannot be predicted any more. Therefore, we stop to solve the balance equations of the RVE for values $\s > 0.6$ and constrain the fluctuations $\tilde{\vec{w}} = \vec{0}$ following the approach in \eqref{eq:Con}$_1$. Note, that the tangential stiffness matrix with regard to \eqref{eq:delta_w}  can be adapted easily by removing the linearization with respect to the fluctuations.

In the first (left) figure of  \ref{fig:macro_RVE_stress}, the von Mises stress distribution of a typical RVE using periodic boundary conditions is displayed. To be specific, we display the RVE with the maximum of the deformation gradient using $\op{max}(\|\vec{F}-\vec{I}\|)$ within the area of the macro system with a phase-field below the threshold parameter $\s = 0.6$. In the second (right) figure, the von Mises distribution of the RVE is displayed, searching again for the maximum of the Froebenius norm of the deformation gradient within the area above the threshold parameter; hence without local fluctuations. Remark, that the volumetric average values of the stresses and the driving force of the phase-field are transferred to the macro-scale and hence, local peak values focused on a small area or volume do not contribute significantly.

\begin{figure}[H]
\begin{center} 
\includegraphics[width=0.785\textwidth]{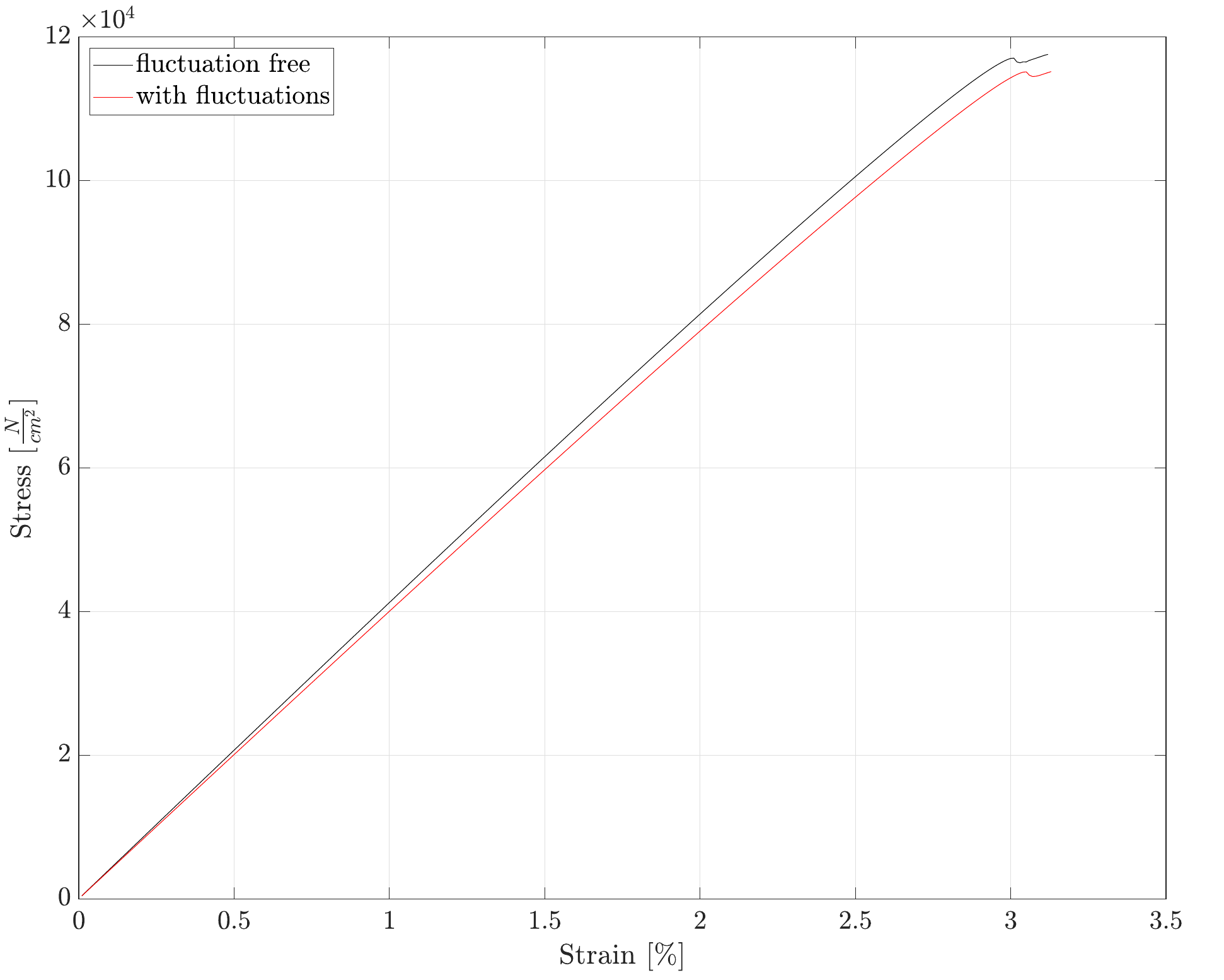} 
\caption{Macro-fracture: Load deflection diagram for the pure Macro system without dissipation at the micro-scale. The RVEs are calculated with and without fluctuations.}\label{fig:macroF_forceDisp}
\end{center}
\end{figure}

Finally, figure \ref{fig:macroF_forceDisp} displays the load–deflection curve, compared to results excluding local fluctuations, i.e. using a Voight type homogenization as shown in \eqref{eq:Con}$_1$. As can be seen, calculation of local deformations reduces the arising, averaged driving forces of the phase-field on the macro-scale. Consequently, fracture initiates at larger displacements.%, as shown in \ref{fig:macroF_forceDisp}.

%% file: sections/conclusions.tex
\section{Conclusions}\label{sec:con}
Phase-field fracture approaches are now used for more than a decade with tremendous success, mainly due to the consistent variational structure and its easy applicability in complex situations. It is most natural to adapt this approach on mutliscale analysis. Two different phase-fields are applied, the first on the micro- and the second on the macroscale.  For a thermodynamically consistent formulation, only one of both phase-fields can be active at the same time,  although a joint framework has been presented.

The first micro phase-field allows for local fracture in the microstructure, using a phenomenological, non-linear strain energy function with a split of the tensile stretches.  This leads to an anisotropic  degeneration of the macroscale stress tensor, calculated as the volumetric average of the microscale stresses.  Hence, from a macroscale point of view this approach acts like an anisotropic damage formulation.  As shown, a full fracture of the RVE using periodic boundary conditions, i.e.\ an infinite crack through the microscale leads to local instabilities, as the required scale separation does not hold any more.

On the other hand, the macroscale phase-field approach suffers not from any instabilities, as usual for non-local gradient formulations.  Therefore, we calculate the phase-field driving force analogues to the stress tensor using a volumetric average on the RVE. Within a joint framework, we can show that this formulation can be considered as a multiplicative triple split of the tensile stretches, resulting in an additive decomposition of the different dissipation terms, the micro- and the macro-dissipation.

\section*{Acknowledgments}
Support for this research was provided by the Deutsche Forschungsgemeinschaft (DFG), Germany under grant HE5943/26-1. This support is gratefully acknowledged.  We also like to thank Ralf M\"uller for fruitful discussions on computational homogenization of micro-fracture.